\documentclass[11pt,graphicx,amsmath]{article}
\usepackage{amsmath}
\usepackage{graphicx}
\usepackage{bm}
\usepackage{color}
\usepackage{amssymb}
\usepackage{amsfonts}
\usepackage{comment}
\usepackage{cite}
\usepackage{todonotes}
\usepackage{caption}
\usepackage{subcaption}
\usepackage{appendix}
\usepackage{epstopdf}

\def\be{\begin{equation}}
\def\ee{\end{equation}}
\def\nn{\nonumber}

\def\ba{\begin{eqnarray}}
\def\ea{\end{eqnarray}}
\def\bl#1\el{\begin{align}#1\end{align}}

\def\la{\langle}
\def\ra{\rangle}

\def\be{\begin{equation}}
\def\ee{\end{equation}}
\def\ba{\begin{eqnarray}}
\def\ea{\end{eqnarray}}
\def\nn{\nonumber}
\def\bl#1\el{\begin{align}#1\end{align}}

\title{  The nonlinear equation of correlation function of galaxies
          in the expanding universe
          and the solution in linear approximation }
\author{\small
            \,  Yang  Zhang\thanks{yzh@ustc.edu.cn} , \,
             Bichu Li  \thanks{libichu@mail.ustc.edu.cn}               \\
 \small  Department of  Astronomy,
   CAS Key Laboratory for Researches in Galaxies and Cosmology, \\
 \small  University of Science and Technology of China, Hefei, Anhui, 230026, China \\
 }

 \date{}

\begin{document}

\maketitle

\begin{abstract}

\Large

We present an analytic study of the  density fluctuation
of a Newtonian self-gravity fluid
in the expanding universe with $\Omega_\Lambda+\Omega_m=1$,
which extends our previous work in the static case.
By the use of field theory techniques,
we obtain the nonlinear,  hyperbolic equation of two-point  correlation function $\xi$
of perturbation.
Under the Zel'dolvich approximation
the equation  becomes an integro-differential  equation
and contains also the three-point  and four-point  correlation functions.
By adopting the Groth-Peebles  and Fry-Peebles ansatz,
the equation becomes closed,
contains a pressure term and a delta source term
which were neglected in Davis and Peebles' milestone work.
The equation has three parameters of fluid;
the particle mass $m$ in the source,
the overdensity $\gamma$, and the sound speed $c_s$.
We solve only the linear equation in linear approximation
and apply to the system of galaxies.
We assume two models of $c_s$,
and take an initial power spectrum at a redshift $z=7$,
which  inherits the  relevant imprint from the spectrum of
baryon acoustic oscillations  at  the decoupling.
The solution $\xi({\bf r}, z)$ is growing during expansion,
and contains  $100$Mpc periodic bumps at large scales,
and a main mountain (a global maximum with  $\xi \propto r^{-1}$)
at small scales $r\lesssim 50$Mpc.
The profile of $\xi$ agrees with the observed ones from galaxy and quasar surveys.
The bump separation is given by  the Jeans length $\lambda_J$,
and is  also modified by  $\gamma$ and $c_s$.
Using  a decomposition
we find that the main mountain
is largely generated by the inhomogeneous solution with the source,
and the periodic bumps come from the homogeneous solution with the initial spectrum.
$\lambda_J$ is identified as the correlation scale of the system of galaxies,
distinguished from the clustering scale determined by  $m$.
The corresponding power spectrum has a main peak located around
$k\sim \frac{2\pi}{\lambda_J}$
associated with the  periodic bumps of $\xi$,
and also contains multiwiggles  at high  $k >k_J $
which are developing during evolution even if the initial spectrum has no wiggles.
Since the outcome is affected by
the initial condition and the parameters as well,
it is hard to infer the imprint of baryon acoustic oscillations accurately.
The difficulties with the sound horizon as a distance ruler are pointed out.

\end{abstract}

\

          and classical fields in curved spacetime

98.65.Dx  Large-scale structure of the Universe

98.80.Jk Mathematical  aspects of cosmology

\large

\section{Introduction }

Research  on the large scale structure have achieved great progress,
especially on the observational side.
Large  surveys for galaxies,
such as 6dFGS \cite{Beutler6dFGS2011},
SDSS  \cite{Anderson2014,SanchezSDSS2017},
WiggleZ \cite{Blake2011,RuggeriBlake2020 , KazinWiggleZ2014},
and  surveys for quasars via Ly$\alpha$,
such as  SDSS BOSS \cite{Busca2013,Slosar2013,BausticaBusca2017,Ata2018},
have provided much of the  data of galaxies and quasar.
So far theoretical studies mostly rely on numerical computations and simulations.
Although there have been analytic studies of the density perturbation
in various models,
the comparison between the theoretical density perturbation with
the observational data from surveys is not straightforward.
This is because the density perturbation is a stochastic field,
whereas the system of galaxies in the Universe
can be regarded as a realization of some statistical ensemble.
One needs certain ensemble-averaged quantities,
 such as the correlation functions of the density perturbation,
to make  a comparison.
The correlation functions contain information of both dynamics and statistics.
Even if the  solution of  the density perturbation is known,
to transfer it into the corresponding correlation functions is not easy,
because the probability function of the stochastic process
is not known sufficiently.
We  only know in  general terms
that the pertinent statistic is non-Gaussian,
due to interaction of long-range gravity.
In this regard, the field equation of the correlation function
has  priority over  that of density perturbation,
and is indispensable for a direct comparison of observations of galaxies.

Davis and Peebles \cite{DaviesPeebles1977}
started  with the Liouville's equation of probability function,
and derived a set of  BBGKY  (Bogoliubov-Born-Green-Kirkwood-Yvon)
  equations
of correlation functions and velocity dispersions of galaxies.
In this scheme,  a galaxy is regarded as a point mass
(galaxies are interacting with gravity)
and the system of galaxies
is a many-body  system which is described by some probability distribution.
The BBGKY approach is one of standard methods used  to describe
the dynamics and evolution of many-body systems.
In this approach, multiple moments of various orders are involved
with multiple variables, and each moment has an equation.
One performs a cutoff at  a  certain order
by dropping all higher-order  moments,
and works only with the   several remaining equations,
prescribing  appropriate initial conditions for them.
This scheme has worked  well for systems,
such as CMB anisotropies and polarization,
where only the  first several moments are retained
and the outcome still has a high accuracy.
However, for the system of galaxies with  long-range gravity interaction,
this approach may not be so simple.
Higher-order  terms can be important at small scales
and may not simply be ignored.
The remaining equations are often coupled and nonlinear.
Adequate initial conditions may not be easy to specify consistently
for the set of coupled equations
as one usually does not have sufficient information for them at an early stage.
In Ref.\cite{DaviesPeebles1977},
Davis and Peebles arrived at a set of five coupled equations
for five unknown functions,
including the two-point correlation function  $\xi$.
To solve  these equations,
consistent initial conditions  are required for the five unknown functions.
In particular, the equation of $\xi$ is not closed,
and contains two unknown velocity-dispersion functions
which in turn are described by two other  equations.
Even if the initial conditions were given,
the set of coupled partial differential nonlinear equations
will still not be  easy to solve.
Due to these points,  the BBGKY equations have not been fully
applied in  practical studies  of the system of galaxies.

Reference \cite{Saslaw1985}
 used the model of self-gravitating fluid  in thermal  quasiequilibrium
to describe the system of galaxies in the expanding Universe.
Macroscopic thermodynamic variables such as internal energy, entropy, pressure, etc,
were employed to study large-scale structures,
and a  power-law correlation function
was employed to calculate modifications to energy and pressure.
Reference \cite{Saslaw1985} also studied
the BBGKY equations for a single-particle distribution function.
Reference \cite{deVega1996a}
also adopted the model of self-gravitating gas in thermal equilibrium
and the grand partition function.
They focused on the gravitational  potential, instead of the density perturbation,
to study a possible  fractal structure in the space distribution of galaxies.
In these studies  the  equation of correlation function of density perturbation
 has not been given.

In our previous work of the static case
\cite{Zhang2007,ZhangMiao2009},
starting with the hydrodynamic equations of self-gravity fluid,
using the functional derivative method,
we derived the closed, static equation of $\xi$,
and obtained the solution  $\xi(r) \propto r^{-1.7}$
at small scales (with the amplitude being proportional to the mass of galaxy).
At large scales  the solution $\xi(r)$ contains
periodic bumps with a separation $\sim 100$Mpc
which is identified as  the Jeans length.
When applied  to the system of clusters,
$\xi(r)$ also exhibits the scaling behavior \cite{Bahcal1996}.
In this paper, we  study the  case in the expanding Universe.
Using a   similar method,
we   derive the nonlinear, partial equation of the correlation function  $\xi(r,t)$,
which is also an  integro-differential equation.
We obtain its solution in the linear approximation
and apply it to the system of galaxies.
The solution extends the static solution
and provides an account of  the evolution of correlation.
Besides,
it also  distinguishes  the local clustering from the large scale structure,
and reveals the influence of initial conditions.

Section  2 presents
the  nonlinear field equation of  $\xi(r,t)$,
and compares it  with Davis-Peebles' equations.

Section  3 studies the linear equation as an approximation,
 and introduces two working models
for the sound speed of the system of galaxies.

Section  4 presents the ranges of parameters
and the initial power spectrum at a redshift $z=7$
which inherits the imprint of baryon-acoustic oscillations (BAO) at the decoupling.

Section  5 gives the solution
and compares with the observed correlation function
from surveys of galaxies and quasars.

Section  6 analyzes
the periodic bumps in $\xi$
 and the multi  wiggles in the power spectrum.
The difficulties of the sound horizon as a distance ruler are analyzed.

Section 7 analyzes the impact on the solution
from
the expansion,
the sound speed models,
the parameters,
and  the initial condition.
In particular,
a decomposition of the solution into
homogeneous and inhomogeneous solutions
is given.

Section  8  gives the conclusion  and discussion.

Appendix A lists gives the detailed derivation of
nonlinear  equation of $\xi$.
Appendix B  expresses the homogeneous and inhomogeneous
 solutions in terms of the Green's function
to  exhibit the wave nature of the  correlation function.
We use the speed of light, $c=1$,
and the Boltzmann constant, $k_B=1$,
unless otherwise specified.

\section{ The  nonlinear  field equation of two-point  correlation function  }

The current  stage of the   expanding Universe
is described by a flat RW (Robertson-Walker) spacetime background,
and the Friedmann equation is
\[
\big(\frac{\dot{a}}{a} \big)^2
  = \frac{8\pi G}{3} \rho_c [a^{-3}\Omega_m +\Omega_\Lambda],
\]
with $\Omega_\Lambda \simeq 0.7$ and $\Omega_m=1-\Omega_\Lambda $.
The background pressure is small and can be neglected in the Friedmann equation,
so is the radiation component.
We use the normalization $a=1/(1+z)$ in this paper.
A Newtonian self-gravity fluid  system in the expanding Universe
is described by the mass density  $\rho$,  the pressure $p$,
the velocity ${\bf v}$, and the gravitational potential $\phi$.
In the present study the baryons  and dark matter
are coupled by gravity and mixed up.
From the set of hydrodynamical equations of  fluid,
\eqref{continuityexpan}, \eqref{Eulerexpan}, and \eqref{Eulerexpan} in Appendix,
we  obtain the  nonlinear field equation of (rescaled) mass density $\psi$
 (see Appendix A for the derivation)
\be\label{eqevpsiJ}
\ddot\psi + 2H \dot \psi
-\frac{c_s^2 }{a^2} \nabla^2 \psi - 4\pi G\rho_0 (t) (\psi^2 -\psi)
   - \frac{1}{a^2} \nabla \psi \cdot\nabla \phi
   -\frac{1}{a^2} \frac{\partial^2}{\partial x^i \partial x^j}(\psi  v^i v^j)
        =0 ,
\ee
where $H=\dot a/a$,
$\rho_0(t)$ is the mean  mass density of the fluid,
$\psi({\bf x},t) \equiv  \rho({\bf x},t) /\rho_0(t)$ is the rescaled, dimensionless density,
$\phi$  is the potential satisfying the  Poisson equation  \eqref{Poissonphi},
$v^i$ is the peculiar velocity of the fluid,
and $c_s $ is the sound speed  of the fluid,
defined by  $c^2_s = \delta p/ \delta \rho$,
 and is generally  time dependent during the cosmic expansion.
Equation  \eqref{eqevpsiJ} is equivalent to
Eq.(9.19) in Ref.\cite{Peebles1980},
and describes the density of Newtonian self-gravity fluid
in  the expanding Universe,
and holds  for scales inside the horizon of the universe.
For the dust model in a  static universe, $\ddot\psi=0= \dot\psi$ and $v^i=0$,
Eq.\eqref{eqevpsiJ} reduces to
the static equation studied
in Refs.\cite{Zhang2007,ZhangMiao2009}.
(To describe a relativistic fluid in the expanding Universe
one can work with the nonlinear cosmological perturbations
within the framework of general relativity;
see Ref.\cite{WangZhang2017}.)

In the context of cosmology,
the density field $\psi({\bf x}, t)$ is a stochastic field
on the three-dimensional space.
As mentioned in the Introduction, one does not
directly compare $\psi$ with observational data from surveys of the galaxies;
instead,  one computes the theoretical correlation function of $\psi$
in a prescribed statistic, and compares it with
the observed correlation function of the galaxies from the data.
So we seek the equation of the correlation function of  $\psi$
that bears  more direct relevance to observations than the equation of $\psi$ itself.
Unlike Davis-Peebles' scheme working with a many-body system
of  galaxies,
we work with $\psi$ as a continuous field
and employ  techniques in field theory,
where the equation of the two-point  correlation function
of a field can be  routinely derived.
For the density field  $\psi$,
the two-point   connected correlation function is defined as
\[
G^{(2)}\left(\mathbf{x}_{1}- \mathbf{x}_{2}, t \right)
 \equiv\left\langle\delta \psi\left(\mathbf{x}_{1}, t \right)
     \delta \psi\left(\mathbf{x}_{2}, t \right)\right\rangle
\]
where $\delta\psi$ is the perturbation of $\psi$,
and $\left\langle ...\right\rangle$ denotes  the ensemble average
prescribed by \eqref{ZJdef} and \eqref{avergdef} in Appendix A.
Following a standard method in field theory   \cite{Goldenfeld1992},
we derive the equation of $G^{(2)}$.
An   external source  $J(\bf x)$ which is $t$-independent,
is added  to Eq.(\ref{eqevpsiJ}),
and then  we take the ensemble average,
and  apply functional derivative $\frac{\delta}{a^3 \beta\delta J(\textbf{x})}$
to each term,  and then we  set $J=0$.
[For the detailed calculations,
 see from Eq.\eqref{exppsieq}  to Eq.\eqref{fctdr} in  Appendix A.]
We arrive at the  equation of $ G^{(2)}$ as the  following
\bl       \label{eq2ptcorr}
 &    \ddot G^{(2)}(\textbf{x}-\textbf{x}' , t)
    +2  H \dot  G^{(2)}(\textbf{x}-\textbf{x}' , t)
    -\frac{c_s^2 }{a(t)^2}  \nabla^2  G^{(2)}(\textbf{x}-\textbf{x}' , t)
     - 4 \pi G   \rho_0(t)   G^{(2)}(\textbf{x}-\textbf{x}' , t )
                  \nonumber \\
&    + G \rho_0 (t)
   \int  \nabla \cdot \Big(  G^{(3)}(\textbf{x},\textbf{x}',\textbf{x}^{''} ; t)
       \cdot  \nabla  \frac{1}{|\bf x-x''|}    \Big) d^3{\bf x''}
          \nn \\
& -  \frac{1}{a^2} \frac{\partial^2}{\partial x^i \partial x^j}
       \frac{\delta}{ ( a^3 \beta)\delta J(\textbf{x}^{\prime})}
        \langle \psi (\textbf{x}) v^i (\textbf{x}) v^j (\textbf{x}) \rangle \Big|_{J=0}
    = \frac{4\pi G m}{a^3 }   \delta^{(3)}(\bf{x}-\bf{x}') ,
\el
where $m$ is the particle mass of fluid
and  $\beta \equiv  1/4\pi G m$.
For the system of galaxies under study, $m$ is the mass of a typical galaxy.
The Dirac delta function $ \delta^{(3)}(\bf{x}-\bf{x}')$ is independent of time.
So far Eq.\eqref{eq2ptcorr} is exact.
It can be compared with Eq.(47) of Ref.\cite{DaviesPeebles1977}
where the pressure term $c_s^2 \nabla^2 \xi$ was neglected.
Equation  \eqref{eq2ptcorr} still contains
the velocity-dispersion term $ \psi v^i v^j$.
To proceed further,
we express the velocity in terms of the density perturbation
under the Zel'dovich approximation \eqref{Zeldovichapprox}.
After some calculation,
Eq.\eqref{eq2ptcorr} becomes the following:
\bl       \label{eq2ptcorr1}
 &    \ddot G^{(2)}(\textbf{x}-\textbf{x}' , t)
    +2  H \dot  G^{(2)}(\textbf{x}-\textbf{x}' , t)
    -\frac{c_s^2 }{a(t)^2}  \nabla^2  G^{(2)}(\textbf{x}-\textbf{x}' , t)
    - 4 \pi G   \rho_0(t)   G^{(2)}(\textbf{x}-\textbf{x}' , t )
                  \nonumber \\
&    + G \rho_0 (t)
   \int  \nabla \cdot \Big(  G^{(3)}(\textbf{x},\textbf{x}',\textbf{x}^{''})
       \cdot  \nabla  \frac{1}{|\bf x-x''|}    \Big) d^3{\bf x''}
          \nn \\
& -  \frac{H^2  f^2(\Omega_m)}{16 \pi^2}
  \frac{\partial^2}{\partial x^i \partial x^j}
       \iint d^3 y \, d^3 z
   \frac{y^i-x^i}{| {\bf y}-{\bf x}| ^3 }
    \frac{z^j-x^j}{|{\bf z}-{\bf x} | ^3 }
    \bigg ( G^{(2)}({\bf x-x'},t)G^{(2)}({\bf y-z},t)
    \nonumber \\
&       +  G^{(3)}({\bf y, z, x'};t)  + G^{(4)}({\bf x, y, z, x'};t) \bigg)
       = \frac{4\pi G m}{a^3 }   \delta^{(3)}(\bf{x}-\bf{x}') ,
\el
which contains the three-point   and four-point  correlation functions,
$G^{(3)}$ and $G^{(4)}$.
This hierarchy is expected for  a many-body system with interaction,
as well as for a field theory with interaction.
Equation \eqref{eq2ptcorr1} is accurate up to
a numerical factor of the term $G^{(4)}$ in the double integration,
caused by  the Zel'dovich approximation.
To make  Eq.\eqref{eq2ptcorr1} closed for $G^{(2)}$ ,
we adopt the Kirkwood-Groth-Peebles ansatz \cite{Kirkwood1932,GrothPeebles1977}
to  $G^{(3)}$,
and the Fry-Peebles ansatz \cite{FryPeebles1978} to  $G^{(4)}$.
Then Eq.\eqref{eq2ptcorr1} becomes  closed as  follows: 
\bl       \label{eq2ptcorr34}
 &    \ddot \xi (\textbf{x}-\textbf{x}' , t)
    +2  H \dot  \xi (\textbf{x}-\textbf{x}' , t)
    -\frac{c_s^2 }{a(t)^2}  \nabla^2 \xi (\textbf{x}-\textbf{x}' , t)
    - 4 \pi G  \rho_0(t)  \xi (\textbf{x}-\textbf{x}' , t )
                  \nonumber \\
&    + G \rho_0 (t)
   \int  \nabla \cdot \Big(
   Q \Big[ \xi (\mathbf{x, x'}) \xi (\mathbf{x',y})
      + \xi (\mathbf{x',y}) \xi (\mathbf{y,x})
      + \xi (\mathbf{y,x})  \xi (\mathbf{x,x'}) \Big]
             \cdot  \nabla  \frac{1}{|\bf x-y|}    \Big) d^3 y
          \nn \\
& -  \frac{H^2  f^2(\Omega_m)}{16 \pi^2}
  \frac{\partial^2}{\partial x^i \partial x^j}
       \iint d^3 y \, d^3 z
   \frac{y^i-x^i}{| {\bf y}-{\bf x}| ^3 }
    \frac{z^j-x^j}{|{\bf z}-{\bf x} | ^3 }
    \bigg ( \xi ({\bf x-x'},t) \xi ({\bf y-z},t)
       \nonumber \\
&  +  Q \Big[ \xi (\mathbf{y, z}) \xi (\mathbf{z, x'})
      +  \xi (\mathbf{z, x'}) \xi (\mathbf{ x', y})
      +  \xi (\mathbf{ x', y}) \xi (\mathbf{y, z})  \Big]
      \nn \\
&   +R_a \Big[ \xi (\mathbf{x,y}) \xi (\mathbf{y, z}) \xi (\mathbf{z, x'})
           + ,,,\text{(sym. 12 terms)}  \Big]
           \nn  \\
&  +R_b \Big[ \xi (\mathbf{x, y}) \xi (\mathbf{x, z}) \xi (\mathbf{x, x'})
          + ,,,\text{(sym. 4 terms)} \Big] \bigg)
          \nn \\
&        =  \frac{4 \pi G m}{  a^3}   \delta^{(3)}(\bf{x}-\bf{x}'),
\el
where   $\xi \equiv G^{(2)}$,
and the time variable $t$  is skipped from $\xi$  in the integrations
 for ease of   notation.
The undetermined numerical factor of $G^{(4)}$
can be absorbed into the parameters $R_a$ and $R_b$
due to the Zel'dovich approximation.
Thus, Eq.\eqref{eq2ptcorr34} is accurate
to the order of perturbation as it stands,
and the error would be of the order  $G^{(5)}$
which is neglected in this study.
Equation  \eqref{eq2ptcorr34}  is a hyperbolic,
nonlinear,  differential-integro   equation  of $\xi$,
and is valid on subhorizon scales in an expanding universe.
It can be used to describe the correlation function of the system of galaxies,
or of clusters.
It contains three nonlinearity parameters $Q$, $R_a$ and $R_b$
in the nonlinear terms.
Application of Eq.\eqref{eq2ptcorr34} is nontrivial,
due to the integration terms
that are expected to be important at small scales.
The linear terms of Eq.\eqref{eq2ptcorr34} are simple
and will be dominant at larger scales $r \gtrsim 10$Mpc
where $\xi \ll 1$,  as  observations  indicate.

It is enlightening to compare our Eq.\eqref{eq2ptcorr34}
with Davis-Peebles' result \cite{DaviesPeebles1977},
which consists of a set of five  equations  (71a), (71b), (72), (76), and  (79)
for five  unknowns ($\xi$, $v_1^2$, $A$,  $\Pi$,   $\Sigma$),
where  $v_1^2$ is the proper peculiar velocity dispersion,
$A$ is the rescaled relative peculiar velocity,
$\Pi$ and $\Sigma$ are   velocity dispersions.
Our Eq. \eqref{eq2ptcorr34} is similar to their Eq. (72),
but there  are  several differences  including  the following.
First,  our Eq.\eqref{eq2ptcorr34} contains
the pressure term $\frac{c_s^2 }{a^2} \nabla^2 \xi$,
which is crucial in revealing acoustic oscillations in large scale structures;
this term  was ignored in Eq.(72) of Ref.\cite{DaviesPeebles1977}
 as they considered a pressureless gas.
Second,   our Eq.\eqref{eq2ptcorr34}   contains the $\delta^{(3)}$ source term,
which is  standard for an equation of two-point  correlation function.
The   $\delta^{(3)}$ term was dropped in a massless limit
in Eq.(72) of ref.\cite{DaviesPeebles1977}.
As we shall demonstrate, the source term is indispensable,
governs the local clustering at small scale,
and predicts the dependence of the clustering amplitude
 upon the mass of the  galaxy.
Third, our Eq.\eqref{eq2ptcorr34} is closed
for the two-point    correlation function $\xi$,
whereas Eq.(72) of Ref.\cite{DaviesPeebles1977}
still contains two  unknowns ($\Pi$ and $\Sigma$)
since  the  Zel'dovich approximation was not used.

The statistics   of the system of galaxies is  non-Gaussian,
and the two-point   correlation function does not exhaust the statistical
  information of the system.
One may go farther to higher-order correlation functions such as $G^{(3)}$, etc.
Using   similar procedures,  we can get the nonlinear equation of  $G^{(3)}$
which will  contain terms like  $G^{(4)}$ and  $G^{(5)}$ etc.
See Refs.\cite{ZhangMiao2009}
for a simple case of the  static, linear equation of  $G^{(3)}$.
Ideally,
when the solutions of all the correlation functions are obtained,
they would constitute  a complete  description
of the system of galaxies.
In this paper we work only with $G^{(2)}$.

\section{ The linear equation of two-point  correlation function of galaxies }

The full content of Eq.\eqref{eq2ptcorr34}
is complex, and its solution will involve much computation.
In the following  we work only with its linear approximation.
Dropping the nonlinear $\xi^2$ and $\xi^3$ terms, Eq.\eqref{eq2ptcorr34} reduces to
\be\label{linapprg}
  \ddot \xi ({\bf x},t)
    +2  H \dot  \xi ({\bf x},t)
    -\frac{c_s^2 }{a(t)^2}  \nabla^2  \xi ({\bf x},t)
    - 4 \pi G  \rho_0(t)  \xi ({\bf x},t)
 =  \frac{4 \pi G m}{ a^3} \delta^{(3)}({\bf x}) ,
\ee
which is a linear,  hyperbolic equation with a delta source,
a gravity term, and an expansion term,
-all having time-dependent coefficients.
It will give a description of the correlation function at large scales,
-the dropped nonlinear terms would affect the correlation function only at small scales,
-as  the static nonlinear solution indicates \cite{ZhangMiao2009}.
When the time-derivative terms are dropped,
Eq.\eqref{linapprg} reduces to
the static linear equation that was studied in Ref.\cite{Zhang2007}.
We shall apply Eq.\eqref{linapprg}
to the system of galaxies in the expanding Universe,
and $\xi$ is regarded as the correlation function of galaxies,
 $m$  as the mass of typical galaxy,
and  $c_s$  as
 the sound speed of acoustic waves of  the system of galaxies.
The  mean mass density of the fluid can be written as
\[
\rho_0(t)  =    \rho_0(t_0)  a^{-3}(t) ,
\]
where $\rho_0(t_0)$  is the present mean density of the fluid
and  can be written as
\be \label{defgamma}
 \rho_0(t_0)= \gamma  \rho_c \Omega_m
\ee
where $\rho_c$ is the critical density,
and $\gamma$ is the overdensity parameter;
$\gamma \geq 1$, since  the fluid density is generally higher than
the cosmic background density $\rho_c \Omega_m$.
This will take into  account  the fact that the density of the surveyed regions
is generally higher than that  of the cosmic background.
We assume $c_s$ is of  the same magnitude as
the peculiar velocity $v$ of galaxies, $c_s \sim v \sim 10^{-3} c$.
This is analogous to the sound speed in a gas of molecules
which is  the order of the  random velocity of atoms.
However,  the magnitude of  $c_s$ here
is much higher than the sound speed  ($\sim 10^{-6} c$) in a gas of molecules.
The former is determined by gravitational potential between galaxies,
$c_s\sim v\sim  (\frac{G m}{r})^{1/2}$,
whereas the latter is mediated by collision between  molecules.
According to  current cosmology,
a component of  dark matter should  also coexist with galaxies.
Although dark matter is collisionless,
it is coupled with galaxies through gravity,
and  therefore, it should have the same $c_s$ as for the galaxies.
By the energy conservation equation,
the peculiar velocity of galaxies is decreasing in the expanding Universe,
$v \propto a^{-1}(t)$ when galaxies are regarded as point particles,
or   $v \propto a(t)^{-3/5}$
when the two rotational degrees of freedom of the galaxy  are included.
(The circular speed of spiral galaxies is $\sim 10^{-3} c$,
roughly equal to the translational peculiar velocity \cite{BinneyTremaine1987}.)
So  the sound speed can be written as
\bl \label{csat}
 c_s  & = c_{s0} \, a(t)^{-  \eta } ,
\el
where  $c_{s0}$ is the present  value  at $z=0$,
and  $\eta=\frac35 $ when the rotation of galaxy is included,
or $\eta =1$  without galaxy  rotation.
Then,    \eqref{linapprg} is written as
\bl \label{linapprgtcs0}
\ddot \xi ({\bf x},t)
    +2  H \dot  \xi ({\bf x},t)
    - \frac{ c_{s0}^2}{a^{2+2\eta   }(t)}  \nabla^2  \xi ({\bf x},t)
    - \frac{4 \pi G     \gamma  \rho_c \Omega_m }{a^{3}(t)} \,   \xi ({\bf x},t)
&  =  \frac{4 \pi G m}{ a^3 } \delta^{(3)}({\bf x}).
\el
As we shall see,
 even the  linear equation \eqref{linapprgtcs0}
will reveal rich content of
the correlation function of galaxies.
The left-hand side of  Eq. \eqref{linapprgtcs0}
is similar to the equation of density perturbation.
The pressure term $- c_s^2  \nabla^2 \xi$
gives rise to small-scale  acoustic oscillations in  the fluid,
and its role is against the clustering.
The   gravity term $- 4 \pi G  \gamma \rho_0 \xi$
is  the main driving force  for clustering of density perturbations.
The term $2 H \dot \xi$ in \eqref{linapprgtcs0}
is due to the expansion of the Universe,
and has the  effect of suppressing  the growth of clustering.
The inhomogeneous term $4 \pi G m \delta^{(3)}({\bf x})$
is a source for the correlation function,
as in the static case \cite{Zhang2007}.
Its  magnitude   is proportional to $m$,
so that galaxies of higher mass  acquire a higher-clustering amplitude.
When the two time-derivative terms are dropped,
Eq.\eqref{linapprgtcs0}  reduces to  the  static linear equation \cite{Zhang2007}.

Equation  \eqref{linapprgtcs0} can be solved in the $k$-space more conveniently
without specifying the boundary condition.
Using the Fourier transformation,
\bl\label{Fouriertrafm}
\xi ({\bf x},t)=\frac{1}{(2\pi)^{3 }}
   \int d^3k   \,   P_k(t) e^{i \,\bf{k}\cdot\bf{x}}  ,
\el
where $P_k(t)$  is the power spectrum of dimension $[L^3]$,
which is related to  $\Delta^2_k \equiv \frac{k^3}{2\pi^2}  P_k $
often used in the literature.
The Fourier transformation is also written  as
\bl
P_k(t) & =4\pi \int_0^\infty \xi (x,t) \frac{\sin k x}{k x} x^2 \mathrm{d} x ,
       \\
\xi (x,t) & = \frac{1}{2\pi^2} \int_0^\infty P_k(t)
  \frac{\sin k x}{k x} k^2 \mathrm{d} k ,
  \label{Fourierxi}
\el
which are used in concrete computation.
Equation  \eqref{linapprgtcs0} becomes
the  second-order ordinary differential equation of the power spectrum
\bl \label{Jseqkm0}
\frac{d^2 }{dt^2 }   P_k(t)
  + 2H  \frac{d }{dt }    P_k(t)
  +   \frac{ 4\pi G \,   \rho_c \Omega_m  }{a^{2+2\eta }(t)}
     \Big( \frac{k^2}{k^2_{J}} -   \gamma \,  a^{2\eta -1}  (t) \Big)  P_k(t)
 &    = \frac{4\pi G m}{a^3} ,
\el
where
\be \label{kJdef}
k_{J} \equiv  \frac{\sqrt{ 4 \pi G  \,    \rho_c \Omega_m}}{ c_{s0} }
= \big(\frac32 \Omega_m \big)^{1/2} \frac{H_0}{c_{s0}}
\ee
is the present Jeans wave number  (at $z=0$),
and   $\lambda_J= 2\pi/k_J$ is the present  Jeans length
 \cite{Jeans1902,GamowTeller1935,Bonnor1957}.
Note that we use the background density in the definition \eqref{kJdef},
and keep the overdensity $\gamma$   as a separate parameter.
In an expanding universe, by use  of  \eqref{csat},
the Jeans length of the  system of galaxies is actually changing
\bl \label{jeanlengthz}
\lambda_J(t)=
\frac{\sqrt{\pi}c_s}{\sqrt{G\rho_0(t)}}  = a^{\frac32-\eta}  \lambda_J .
\el
Note that $\lambda_J(t)$ generally departures from the
comoving ($\propto a$).
In our paper, $\lambda_J(t) \propto a^{1/2}$ for the model $\eta=1$,
and $\lambda_J(t) \propto a^{9/10}$ for the  model $\eta=\frac35$.
For each fixed  $k$,
Eq.\eqref{Jseqkm0} describes an oscillating mode
 when $k^2/k^2_{J} > \gamma \,  a^{2\eta -1}$,
or a growing mode when  $k^2/k^2_{J} < \gamma \,  a^{2\eta -1}$.
It should be noticed that, during expansion,
more and more oscillating $k$-modes are turning into growing modes
for both  $\eta=1$ and $\eta=\frac35$.
The source $4\pi G m/a^3$ is  $k$-independent,
and appears as an external force acting equally on
all the $k$-modes.
Using  $a$ as the time variable,
Eq.\eqref{Jseqkm0}  is rewritten as
\bl \label{linequavar}
& \frac{\partial^{2}}{\partial a^{2}}  P_k
+ \Big( \frac{3}{a}-\frac{3}{2 a} \frac{a^{-3}
       \Omega_{m}}{ (a^{-3} \Omega_{m}+\Omega_{\Lambda} )} \Big)
           \frac{\partial}{\partial a} P_k
+  \frac{3}{2 a^{4+2\eta }} \frac{  \Omega_{m}}
     {(a^{-3}  \Omega_{m}+\Omega_{\Lambda})}
       \Big( \frac{k^2}{k_{J}^2}  - \gamma\,  a^{2\eta -1}   \Big) P_k
       \nn \\
&  =\frac{A}{a^{5} (a^{-3} \Omega_{m}+\Omega_{\Lambda} )} ,
\el
where the source magnitude
$A \equiv 4\pi Gm/ H_0^2$ is $k$-independent
and proportional to the mass $m$ of a typical galaxy,  or cluster,
 under consideration.
Beside the cosmological parameters $H_0$ and $\Omega_m$,
the evolution equation \eqref{linequavar} contains
$m$,  $k_J$  and  $\gamma$ as three independent parameters of
the self-gravity fluid that models the system of galaxies.
The present sound speed $c_{s0}$ is  absorbed into $k_J$,
and will not be regarded as an independent parameter.

\section{ The initial condition and the parameters }

To solve the differential equation \eqref{linequavar},
an appropriate  initial spectrum is needed.
Currently we do not know
the correlation function of galaxies at the early stages ($z \gtrsim 7$)
from observations. (See \cite{Peebles1993} for a review.)
To be specific for computation,
we adopt an analytic initial power spectrum
\bl\label{iniPkold}
P_{k\, ini}(z) =\frac{1}{2n_0  (1+z)^{3}}\frac{1}{|(\frac{k}{k_0})^2 -a(z)|} ,
\el
where $z=7$ will be taken when galaxies,  or protogalaxies,
have been formed.
Equation \eqref{iniPkold}  is based on an extension of
the analytic solution of the static linear equation  \cite{Zhang2007},
and has a similar profile to the initial linear spectrum
used in simulations \cite{Springel2018}.
The initial correlation function associated with  \eqref{iniPkold}
contains small seeds of bumps distributed over the whole space
(see Fig.\ref{2} and  Fig.\ref{3}),
and is  consistent with the
homogeneity and isotropy of the background spacetime.

\begin{figure}
\centering
\includegraphics[width=0.5 \textwidth]{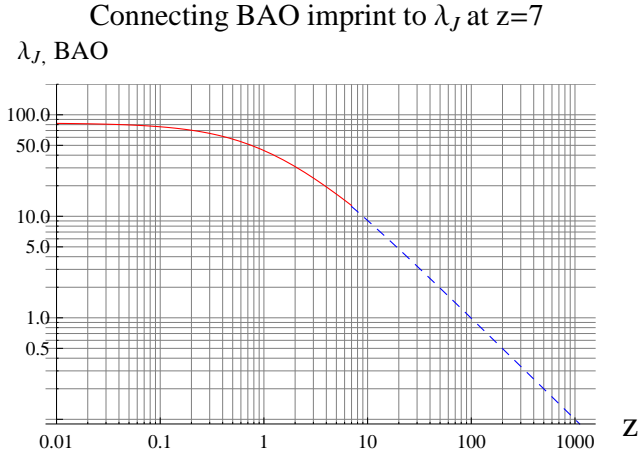}
\caption{
Solid: the Jeans length $\lambda_J(t) \propto a^{0.9}$ for $\eta=\frac35$;
Dashed: the comoving  BAO imprint.
They are connected at $z=7$,
ie, the imprint is transferred to $\lambda_J$  at $z=7$
as the initial condition.
}
\label{1}
\end{figure}

The initial amplitude of  \eqref{iniPkold}
is given by a range
$1/2 n_0   \simeq  (4 \sim 8) \times 10^3  \,  h^3 \, Mpc^{3}$,
where  $n_0$ represents  the  meaning of the  number density  of galaxies.
In  \eqref{iniPkold}  an absolute value is used
for a positive initial spectrum at small $k$,
and  a cutoff of height $(\sim 1000$Mpc$^{3})$
is taken to avoid divergence at $k=k_0$.
The  characteristic wave  number $k_0$  in \eqref{iniPkold} is very important,
as it determines the peak location of the  spectrum  at the initial epoch.
To fit with the observed correlation function
of galaxies \cite{RuggeriBlake2020},
and of quasars in Ref.\cite{Ata2018},
we can take the following range of values
\bl \label{k0ini}
 k_0 \simeq (0.23 \sim 0.57 ) h \,  \text{Mpc}^{-1},
     ~~~ ie,~ \lambda_0 \simeq  (11 \sim  27) \text{Mpc}
     ~~ \text{at $z=7$}.
\el
Within this broad range,
lower  values of $k_0$ are used for  the model $\eta=3/5$,
and higher values of $k_0$ are used the model  $\eta=1$.
The  range \eqref{k0ini}  includes the  imprints
of BAO at the decoupling  ($z\sim 1020$)
\cite{SunyaevZeldovich1970,PeeblesYu1970}
that have survived the Silk damping \cite{Silk1968,Field1971,Weinberg1971,Weinberg1972}
 and comoved up to the epoch $z=7$.
We give a brief illustration.
The BAOs prior to the decoupling are standing waves of baryon-photon plasma
with certain intrinsic wavelengths,
which are allowed to exist above the the scale of
the Silk damping \cite{Silk1968,Field1971,Weinberg1971,Weinberg1972}
and well inside the Hubble radius $1/H$.
Consider the spectrum of BAO at the decoupling
in Fig. 4 of Ref. \cite{PeeblesYu1970}
 for a pure baryon model $\Omega_b \sim 0.03$.
It contains four characteristic peaks which  survive the Silk damping.
The first two have the  comoving wavelengths $230$ Mpc and $540$ Mpc
which are too large, beyond the current observations,
and we do not consider here.
The last two peaks have the  comoving wavelengths
 $100$Mpc and $140 $Mpc  approximately,
and their imprints at $z=7$  are, respectively
\bl \label{impr2}
\lambda  \sim   13  \text{Mpc}, \,    18  \text{Mpc}  ,
\el
which fall into the range \eqref{k0ini}.
When  cold dark matter (CDM) is present,
the values of  BAO imprints in \eqref{impr2} will be modified,
but they will still fall into the range \eqref{k0ini}.
(See also Refs.\cite{HuSugiyama1995,Holtzman1989,EisensteinHu1998}
for the  models with baryons plus CDM.)
After the decoupling,  the imprints of characteristic BAO modes
are influenced by the  gravity of small density fluctuation
and their stretching is generally  a bit slower than  the comoving
in a  model-dependent fashion.
Its detail is  worthy of study in future.
Actually we are not concerned with the precise value
of the  characteristic wavelengths of BAO,
as long as some of them fall into the broad range \eqref{k0ini}.
For illustration,  Fig.\ref{1} shows a connection of
the BAO  imprint $\lambda  \sim 13$ Mpc of \eqref{impr2}
to the Jeans length   of the system of galaxies  at $z=7$.
Thereby, the imprint is transferred to $\lambda_J$ at $z=7$ as the initial condition.
In this  way,  the initial spectrum \eqref{iniPkold}  with  \eqref{k0ini}
of the system of galaxies incorporates a relevant part of the BAO spectrum.
During the evolution from $z=7$ to $z=0$
the Jeans length is relevant which has replaced the BAO imprint,
the final value of Jeans length at $z=0$ is $\simeq 83$ Mpc,
which is  lower than the would-be comoving BAO imprint $\simeq 100$ Mpc
 at $z=0$.
We remark that the choice of  initial spectrum \eqref{iniPkold} is not unique,
and other alternative choices are allowed.

An initial rate is also needed to solve  \eqref{linequavar}.
Define the rate $r_a$ by
\be
\frac{\partial}{\partial a}   P_k (a) =  r_a   P_k (a)  .
\label{ratedef}
\ee
The  conservation of pair  \eqref{cont2pgr2} at linear level
gives
\be  \label{raterv}
 r_a(z) = \frac{f(\Omega_m, z)}{a}= (1+z) \, f(\Omega_m, z),
\ee
where $f(\Omega_m, z)$ is given by Eq.\eqref{fOmega}.
From this,
we get an estimate of the initial rate  $r_a \simeq  7$ at $z=7$.
As it turns out,   the outcomes $\xi$ and $P_k$
are  actually not sensitive to the value of $r_a$
within two orders of magnitude.

 The parameters appearing  in Eq.\eqref{linequavar}
 are given in the following.
The cosmological parameters are taken
in the range  $\Omega_m = (0.25 \sim 0.30)$,
and $H_0^{-1} =3000 \, h^{-1}$ Mpc with $h = (0.69\sim 0.73)$ as default,
and the outcome correlation function
does not change much within the range.
To fit with the observed correlation function of galaxies,
and of quasars,  from surveys \cite{RuggeriBlake2020,Ata2018},
we   take the three parameters of fluid in the following range
\bl
k_J   &  =  ( 0.045  \sim 0.088 ) \,  h\, \mathrm{Mpc}^{-1} ,
    ~~ i.e., ~\lambda_J= (139 \sim 71) \,  h^{-1} \text{Mpc}
     \label{parameters}
     \\
A  & = (2 \sim 5) \times 10^3 \,  h^{-2} \mathrm{Mpc}^3 ,
         \label{parameterA}
         \\
\gamma  &  =   1 \sim  6 .
          \label{gammap}
\el
By \eqref{jeanlengthz},
 $k_J$ and  $k_0$ should be  related   as follows
\bl
k_{0} & \simeq  a^{-\frac32 +\eta} k_J =  (1+z)^{\frac32 - \eta}  k_J  ,
        \label{k0K0}
\el
which is taken only approximately in our  computation.
The range \eqref{parameters} of  $k_J$
approximately  corresponds to the  range   \eqref{k0ini} of  $k_0$.
From these parameters
the sound speed is inferred
\be
c_{s0} = (\frac32 \,  \Omega_m)^{1/2}\frac{ H_0}{k_J}
= (2.3 \sim 4.5) \times 10^{-3}c
 \sim  (0.6\sim 1.2) \times 10^3  (\frac{km}{s}),
\ee
which is  slightly higher than
the observed peculiar velocity of galaxies \cite{Peebles1993}.
The particle mass is inferred  as
\bl\label{mass}
m= \frac{H_0^2  A}{4\pi G}
       \sim   10^{14} M_{\bigodot}  ,
      ~~\text{for $A = 1000$Mpc$^{3}$}
\el
which is larger than a typical galaxy mass $10^{12} M_{\bigodot}$,
and  comparable to that of a cluster.
The inferred $m$ is expected to be reduced
when the nonlinear terms of Eq.\eqref{eq2ptcorr1}
that will enhance  clustering  substantially at small scales
are included.

\section{\bf The linear solution and its comparison with observations}

\begin{figure}
\centering
\includegraphics[width=0.7 \textwidth]{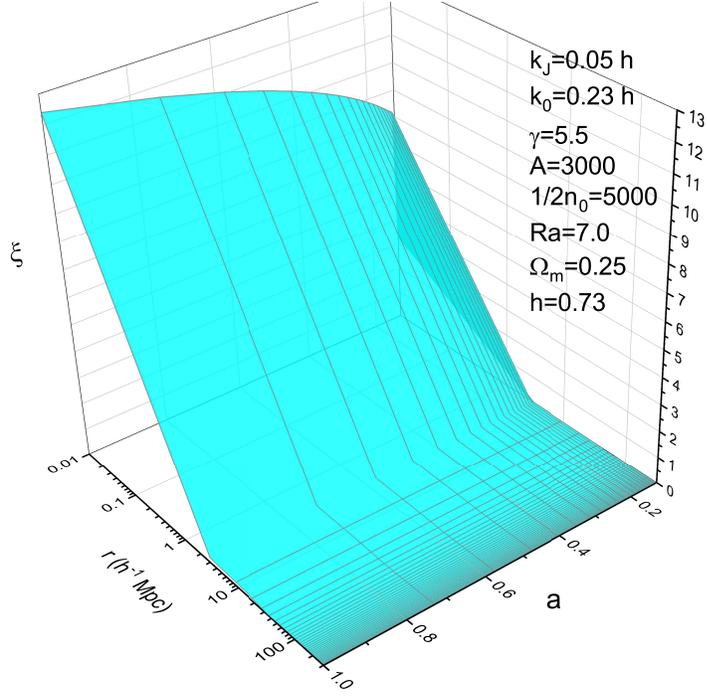}
\caption{ The evolution of correlation function $\xi(r,a)$
 from  $a=1/8$ to $a=1$ in the model $c_s=c_{s0}a^{-3/5}$.}
\label{2}
\end{figure}

\begin{figure}
\centering
\includegraphics[width=0.7 \textwidth]{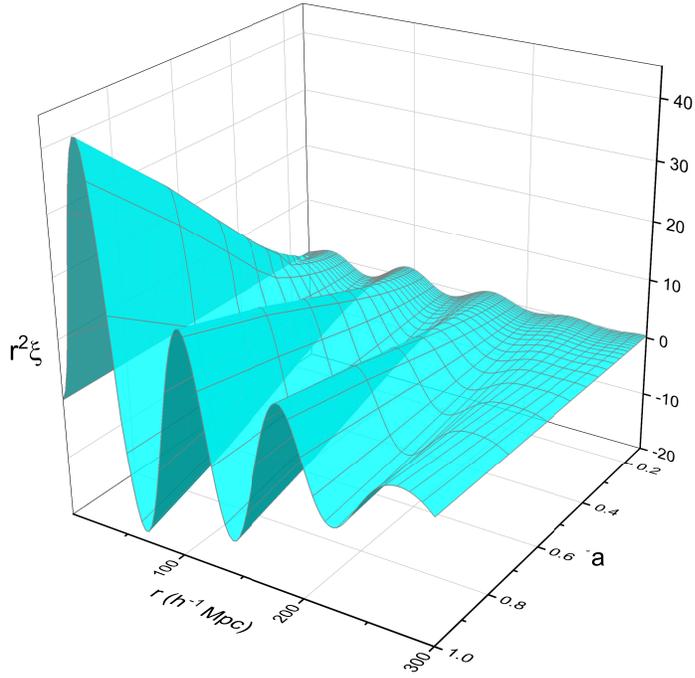}
\caption{ Same as  Fig.\ref{2}.
The weighted $r^2 \xi(r,a)$  exhibits the periodic bumps.}
\label{3}
\end{figure}

\begin{figure}
\centering
\includegraphics[width=0.7 \textwidth]{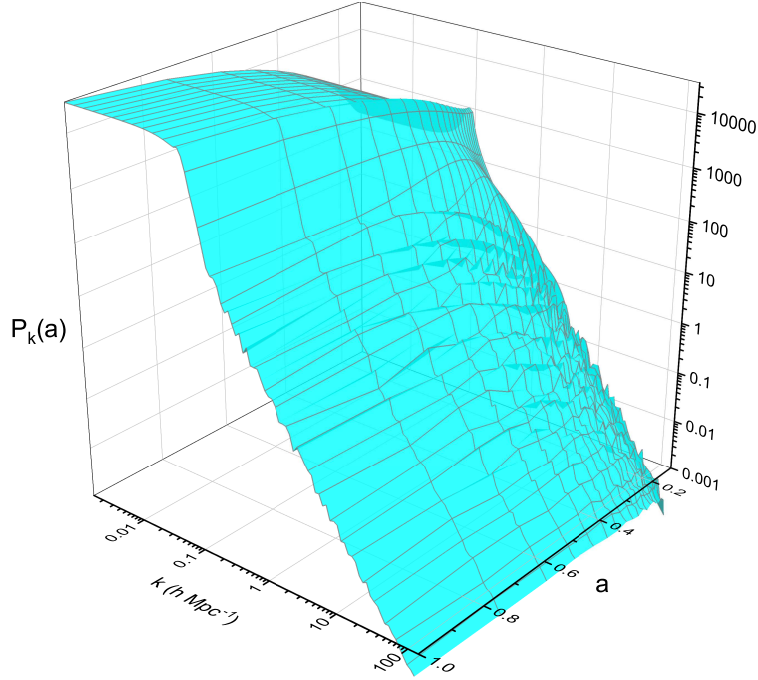}
\caption{ Same as  Fig.\ref{2}.
The evolution of  power spectrum $P_k(a)$.
The main peak is at $k \lesssim k_J$,
and the multi wiggles are developing on the main slope at $k >  k_J$.}
\label{4}
\end{figure}

\begin{figure}
\centering
\includegraphics[width=0.7 \textwidth]{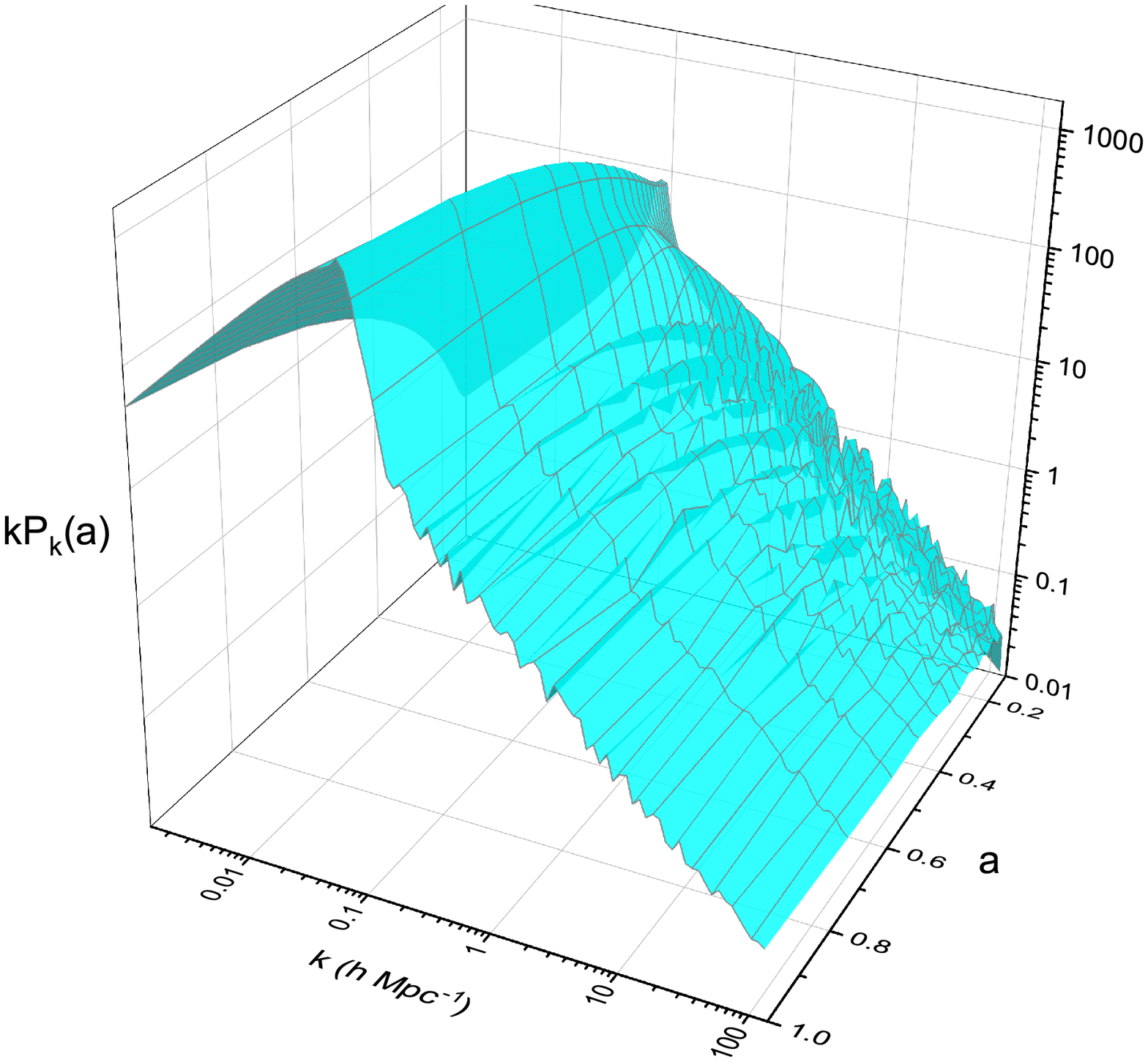}
\caption{Same as  Fig.\ref{4}.
The weighted $k P_k(a)$ exhibits the main peak at $k\sim  k_J$.}
\label{5}
\end{figure}

Given the  initial conditions and parameters,
by solving Eq. \eqref{linequavar} for each $k$ from  $z=7$ to $z= 0$,
we obtain $P_k(z)$ as a function of $(k,z)$,
and by Fourier transformation we also obtain  $\xi(r,z)$ as a function of $(r,z)$.
They are  plotted  in Fig. \ref{2} - Fig. \ref{5},
for the sound speed model  $c_s=c_{s0}a^{-3/5}$.
The  model  $c_s=c_{s0}a^{-1}$ has an analogous outcome,
its two-surface graphs are not shown to save room.

During evolution  from $z=(7\sim 0)$,
the profile of $P_k(z)$ keeps a shape  similar to the initial power spectrum,
increasing in amplitude and developing small wiggles.
For $\xi(r,z)$,
the separation between periodic bump feature
is stretching to a greater distance;
the bumps  are  getting higher  and the troughs are getting lower.
The solution  demonstrates  that  during  $z=(7\sim 0)$
the correlation function at large scales  keeps a similar pattern
and there is no abrupt change.
In this sense we may say that  in the expanding Universe
the distribution of galaxies is in an asymptotically relaxed state
\cite{Saslaw1985}.

\begin{figure}
\centering
\includegraphics[width=0.7\textwidth]{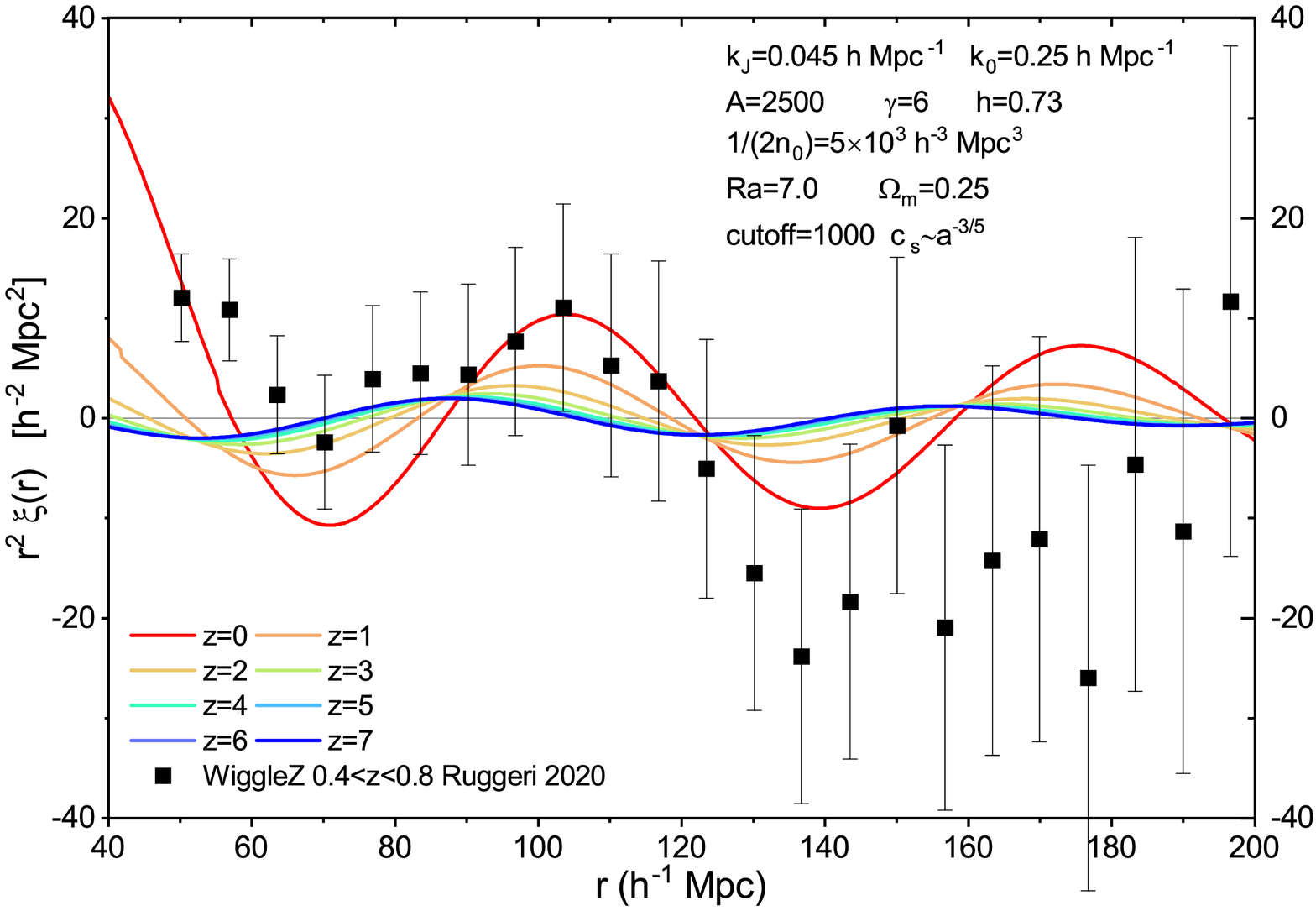}
\caption{ $r^2 \xi$ of the  model $c_s=c_{s0}/a^{3/5}$ compared with
the   data of WiggleZ galaxies
in  Ref. \cite{RuggeriBlake2020}.
The evolution from $z=7$ to $z=0$ are also shown.
}
\label{6}
\end{figure}

\begin{figure}
\centering
\includegraphics[width=0.6 \textwidth]{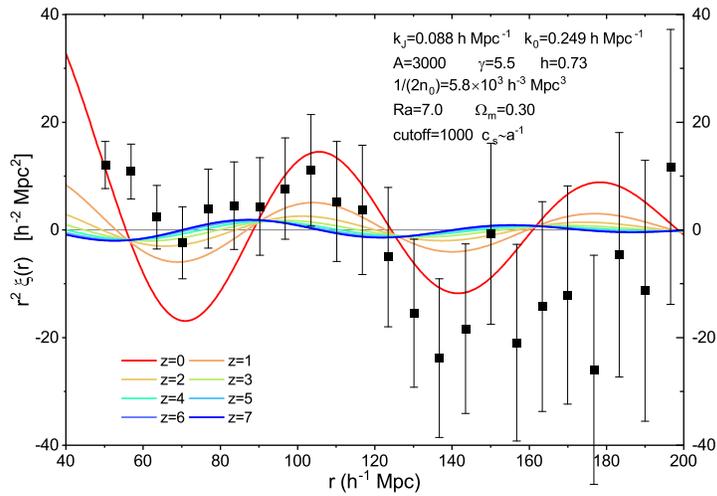}
\caption{  $r^2 \xi$ of the model $c_s=c_{s0}/a$
compared with the   data of WiggleZ galaxies
in  Ref. \cite{RuggeriBlake2020}.
}
\label{7}
\end{figure}

We compare the   solution $\xi(r)$
with the latest observed correlation of WiggleZ galaxies \cite{RuggeriBlake2020}
in Fig. \ref{6} for the model $c_s=c_{s0}/a^{3/5}$,
and in Fig. \ref{7}  for the model $c_s=c_{s0}/a$.

\begin{figure}
\centering
\includegraphics[width=0.6 \textwidth]{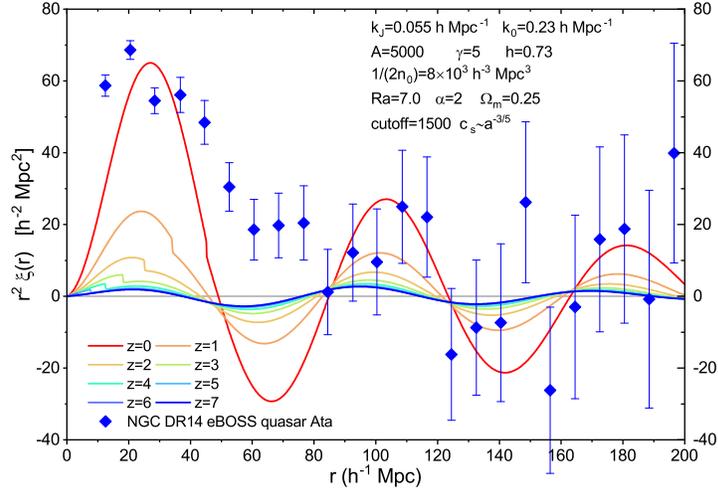}
\caption{ $r^2 \xi$ of the model $c_s=c_{s0}/a^{3/5}$
compared with the  data of NGC quasars in Ref.\cite{Ata2018}.}
\label{8}
\end{figure}

\begin{figure}
\centering
\includegraphics[width=0.6 \textwidth]{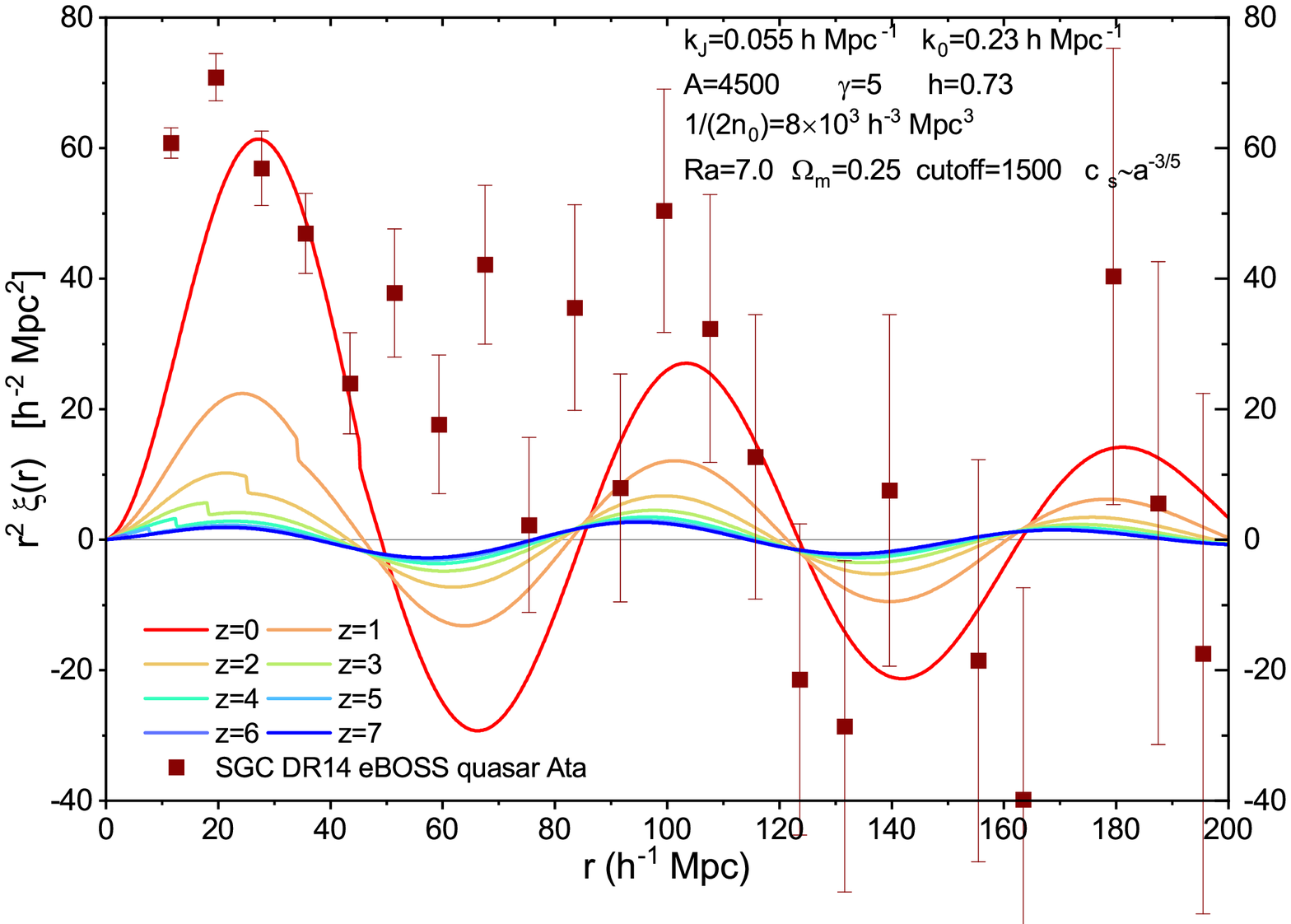}
\caption{ $r^2 \xi$ of the model $c_s=c_{s0}/a^{3/5}$
compared with the   data of SGC quasars in Ref.\cite{Ata2018}.}
\label{9}
\end{figure}

\begin{figure}
\centering
\includegraphics[width=0.6 \textwidth]{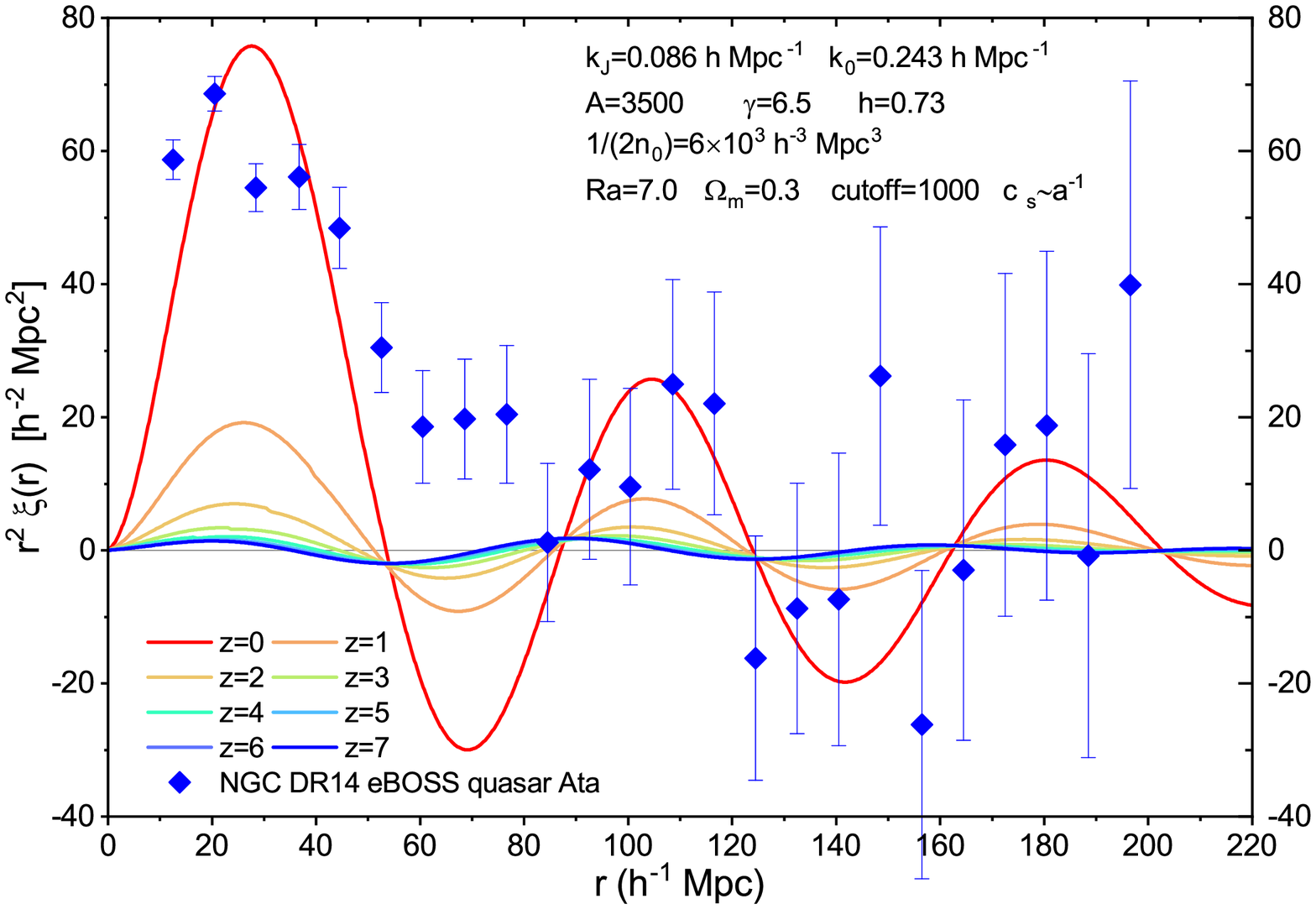}
\caption{ $r^2 \xi$ of the model $c_s=c_{s0}/a$ compared with
the  data of NGC quasars in Ref. \cite{Ata2018}.}
\label{10}
\end{figure}

\begin{figure}
\centering
\includegraphics[width=0.6 \textwidth]{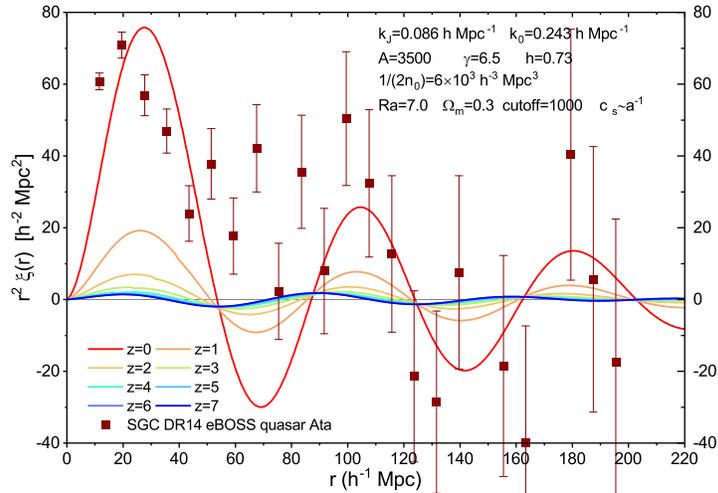}
\caption{ $r^2 \xi$ of the model $c_s=c_{s0}/a$ compared with
the   data of SGC quasars in Ref. \cite{Ata2018}.}
\label{11}
\end{figure}

We also compare the solution
with the observed  NGC and SGC quasar data (Ref. \cite{Ata2018})
in Fig. \ref{8} and  Fig. \ref{9}  for model $c_s=c_{s0}/a^{3/5}$,
and in Fig. \ref{10} and  Fig. \ref{11} for the model $c_s=c_{s0}/a$.
It is seen that the weighted correlation function $r^2 \xi(r)$
possesses   periodic oscillatory bumps along the distance $r$;
the height of bumps and the separation between bumps are  close to the observed ones.
Generally the galaxy and quasar surveys cover regions
with  different physical environments,
so we may choose different values of the parameters
within  the range listed in Eqs. \eqref{parameters}-\eqref{gammap}.


Equation  \eqref{linapprgtcs0} can also apply to
the correlation function of clusters
when the appropriate parameters $m$ and $c_{s0}$ are used for the system of clusters.
The mathematical structure of Eq. \eqref{linapprgtcs0}
remains the same for galaxies and for clusters,
except that clusters have a higher source amplitude $A\propto m$.
This explains why the observed correlation functions of clusters
have  a similar profile to that of galaxies,
but with a  higher amplitude  at small scales.
These properties were previously predicted
in the static case \cite{Zhang2007,ZhangMiao2009},
and also hold  in the  expanding Universe.

\begin{figure}
\centering
\includegraphics[width=0.6\textwidth]{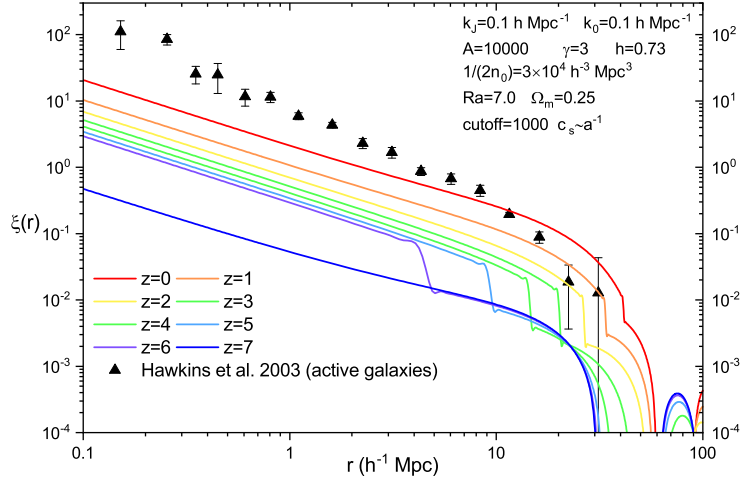}\\
\caption{   The model $c_s=c_{s0}/a$.
The growing  $\xi(r)$ from $z=7$ to $z=0$.
The observational data (triangle) is from Ref. \cite{Hawkins2003}.}
\label{12}
\end{figure}

\begin{figure}
\centering
\includegraphics[width=0.6\textwidth]{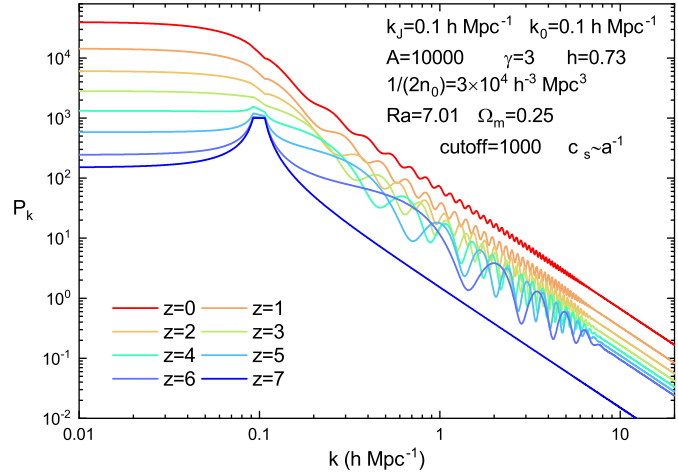}\\
\caption{  Same as   Fig. \ref{12}.
The growing  power spectrum $ P_k$  from $z=7$ to $z=0$.}
\label{13}
\end{figure}

The  amplitudes of  $\xi$ and $P_k$
are increasing  during  evolution  $z=(7\sim 0)$,
as seen in Fig. \ref{12} and Fig. \ref{13} for the model $c_s=c_{s0}a^{-1}$.
The growth varies with different scales.
For instance,  roughly $\xi(a)$, $P_k(a)  \propto a^{0.2} $
at small scales ($\xi$ at  $r=1 h^{-1}$ Mpc,   $P_k$ at $k=1 h$ Mpc$^{-1}$),
and $P_k(a) \propto a^{1.4} $  at large scales  ($k=0.05 h $ Mpc$^{-1}$).

The profile of  the linear solution  $\xi(r)$
has a global maximum with a power-law slope,
 $\xi \propto r^{-1}$ at $r\lesssim 10$Mpc,
which is referred to as the main mountain in this paper.
The slope is flatter than the observed [$\xi = (r/r_0)^{-1.7}$]
  \cite{TotsujiKihara1969,Peebles1974a, Peebles1974b},
and  will not be  improved by merely increasing the mass $m$.
This reflects the limitation of  the linear solution at small scales.
We expect the mountain will get higher and steeper by nonlinear terms,
as has been shown in the static nonlinear  solution \cite{ZhangMiao2009}.
At high $k> k_J$, the profile of spectrum  $P_k \propto k^{-2}$,
the same as  that of the initial power spectrum \eqref{iniPkold}.
This is referred to as the main slope of $P_k$
and corresponds to the main mountain of $\xi$.
Note that the linear spectrum $P_k$ at high $k$
will lead to a divergent  autocorrelation $\xi(0)$  in \eqref{Fourierxi}
at the upper limit $k = \infty$ of integration.
This UV divergent behavior is analogous to
an inflaton scalar field in the inflationary  Universe,
and can be removed by adiabatic regularization to an appropriate order
\cite{ZhangYeWang2019,ZhangWangYe2019}.
If  the nonlinear terms are included,
the behavior of $P_k$ at high $k$ is expected to be modified,
deviating from  the linear one.

\section{The bumps and the wiggles}

We analyze  the prominent features,
i.e., the bumps in $\xi$ and the wiggles in $P_k$,
  in the following.

(1)  The most   prominent  features are
  the periodic bumps in the correlation function.
At large distance,   $\xi$ consists of the periodic bumps
which   are  approximately located at
\bl
r \sim    100,  \, 180,  \, 270 \,  \text{Mpc}, ... ,
\label{bumpsloc}
\el
with  the separation between two neighboring bumps is  approximately
\bl \label{sepbump}
\Delta r \sim  (80\sim 100) \,  \text{Mpc}.
\el
The  bumps are more clearly seen
in the weighted correlation $r^2 \xi(r)$ plotted in
Fig. \ref{3},
Figs.\ref{6}-\ref{11}.
The data from surveys of galaxies \cite{RuggeriBlake2020}
and of quasars  \cite{Ata2018}
exhibit   the first bump at 100 Mpc,
and the negative trough on the interval  ($130 \sim 160$)Mpc,
as well as indicate the existence of a second bump at $\sim 200$Mpc.
This agrees   with the prediction  \eqref{bumpsloc}.
The 100Mpc periodic bumps were predicted
in the static solution \cite{Zhang2007,ZhangMiao2009},
and now also show up  in the evolution solution in the expanding Unverse.
The  bump locations \eqref{bumpsloc}
and the separation $\Delta r$  \eqref{sepbump}
are largely determined by   $\lambda_J$,
but also affected
by the overdensity  $\gamma$, the sound-speed model,
 the cosmic  expansion,
and   the details of the initial spectrum as well,
as we shall analyze  in Sect 7.

Early pencil-beam redshift surveys already showed the 100 Mpc periodic feature
in the correlation function of galaxies
 \cite{Broadhurst1990,Broadhurst1995,Tucker1997}
 and of clusters  \cite{Einasto1997a,Einasto1997b,Einasto2002a,Einasto2002b,Tago2002}.
Recent surveys have already  shown  the existence of two bumps,
one  at $\sim 100$Mpc and another at $\sim 200$ Mpc,
in the correlation function of galaxy
\cite{RuggeriBlake2020,Beutler6dFGS2011,Anderson2014,Anderson2012,KazinWiggleZ2014,SanchezSDSS2017},
as well as  of quasars
\cite{Ata2018,BausticaBusca2017,Busca2013,Blake2011,Agathe2019}.
All these observational  results  confirm
the prediction of a  periodic feature  \eqref{bumpsloc}.
Some simulations also show this phenomenon \cite{Yahata2005,EinastoAA2002}.
Large surveys in the future might have a chance of detecting
  the third bump  at $\sim 300$ Mpc in \eqref{bumpsloc}.

The  100 Mpc periodic bump  feature follows  from  Eq.\eqref{linapprgtcs0}
by a qualitative analysis.
For simplicity,  we let $a=1$ and neglect the subdominant expansion term $2H \dot \xi$
as approximation,
then Eq.\eqref{linapprgtcs0} reduces to
\[
\ddot \xi - c_{s0}^2 \nabla^2 \xi - 4 \pi G   \rho_0 \xi =4\pi G m \delta^{(3)}({\bf r}).
\]
Using  a time-frequency Fourier transformation
$\xi ({\bf r}, t) = \frac{1}{2\pi}
    \int d \omega \, \xi_{\omega} ({\bf r}, \omega) e^{- i\omega t}$,
we can solve a Helmholtz equation for each frequency mode $ \xi_{\omega}$,
and get
\bl \label{lmodes}
\xi(r,t) \sim   \sum_\omega b_{\omega}
  \frac{  G m}{r} \cos  \Big[  (k_J^2  + \frac{\omega^2}{c_{s0}^2} )^{1/2} r \Big]
  e^{-i\omega t}.
\el
The  lowest-frequency mode in \eqref{lmodes}
reduces to the static solution  \cite{Chavanis2006,Zhang2007}
\bl \label{0mode}
\xi  \sim b_0 \frac{G m  \cos (k_J \, r)}{r} ,
\el
where $\cos (k_J \, r)$ gives rise to the periodic bumps
with  the  separation being
the Jeans  length $\lambda_J=2\pi/k_J$.
Besides,  there are other modes in \eqref{lmodes},
oscillating  at various higher frequencies,
referred to as the sub-bumps,
whose wavelengths are roughly  a fraction of the Jeans length,
\bl\label{subbumps}
\lambda \simeq  \frac{\lambda_J}{l}, ~~ l=2,3,4, ...  ,
\el
and their amplitudes   are much lower than that of the bumps,
by orders of   magnitude,  and thus are  barely noticeable in the  graphs.
The  current observational data are insufficient to exhibit these sub-bumps either.
These sub-bumps are associated with the wiggles in $P_k$,
 as we shall analyze later.

Since its discovery in 1990s the $100$ Mpc periodic bump feature
has been interpreted by various tentative models
 \cite{Broadhurst1990,Broadhurst1995,Tucker1997,
Einasto1997a,Einasto1997b,Einasto2002a,Einasto2002b,Tago2002}.
More recently it  was  interpreted as being caused
by the imprint of the sound horizon  \cite{Eisenstein2005}.
In the following we analyze the issue of sound horizon
and clarify certain statistical concepts involved.
The sound horizon was  defined
as an integration from $z=\infty $  to  the decoupling epoch
\cite{HuSugiyama1995,EisensteinHu1998}
\bl \label{soundhorizon1}
s & = \int_0^{t(z_d)} \frac{c_s}{a(t)} \, dt
= \int_0^{t(z_d)}  c_s\, (1+z) \, dt ,
\el
where  the sound speed of baryon  gas  is
\cite{SunyaevZeldovich1970,PeeblesYu1970,HuSugiyama1995,EisensteinHu1998}
\be \label{sspd1}
c_s= \frac{1}{  \sqrt{3(1+R)}},
~~~  R= \frac{3\rho_b}{4\rho_\gamma}
   \simeq  31.5 \Omega_b \, h^2 (\frac{1000}{z}) .
\ee
For  $\Omega_m =0.30$, $\Omega_b=0.045$,  and  $h=0.70$
as the default in this analysis,
the decoupling is  $z_d \simeq 1020$,
 the  integration \eqref{soundhorizon1} gives
\bl\label{sdhorz}
s &  \simeq  167 \,   \text{Mpc} .
\el
For $a=1/(1+z)$,  the value  \eqref{sdhorz} is
also the  the present proper length  \cite{DHWeinberg2013}.
The sound horizon \eqref{soundhorizon1}
is sometimes interpreted as the comoving distance that baryon acoustic sound waves travel.
With  the  observations of correlation function of  galaxies,
the value \eqref{sdhorz}
is not comparable to the observed $100$ Mpc feature, instead, higher  by about  60\%.
Moreover, the sound horizon as a distance ruler
can not give a simple explanation of
the  negative trough at ($130 \sim 160$)Mpc,
nor  the second bump at $\sim 200$Mpc.
(See Figs. \ref{6}-\ref{12}.)
Therefore,  the conventional interpretation of the observed features
in terms of the distance traveled by the BAO  waves
is in doubt and needs to be reexamined.
In Ref. \cite{HuSugiyama1995} on  CMB anisotropies and BAO,
the sound horizon \eqref{soundhorizon1} together with $k$,
occurs as  the phase $\int \omega d\tau = \int k c_s d\tau =ks$
of $k$-modes of BAO,
but not as a distance  that waves travel.
Waves of small density perturbations in the baryons and photons  around the decoupling
can be described by a homogeneous-Gaussian stochastic  process
on the three-dimensional  space
\cite{BardeenBondKaiserSzalay1986,Allen1997,WangZhanglisa2019}.
The important point is
that the path of the wave is unobservable statistically,
as  is the distance of the path.
For instance, in Ref.\cite{BashinskyBertschinger2002},
the plot of the potential  $\phi_r (r,t)$
of the baryon-photon density perturbation in position space
is given, for illustration purpose,
with a fixed normalized amplitude and a fixed initial point.
But the actual situation is not  deterministic
and  $\phi_r (r,t)$ is, in fact,   a Gaussian random field on three-dimensional space.
At a fixed time, the potential $\phi_r (r_i,t)$
is a Gaussian random variable at each point $r_i$,
and  its  $n$-point probability distribution is a multivariate Gaussian,
schematically written as \cite{BardeenBondKaiserSzalay1986}
$P(r_1,...,r_n)=  [(2\pi)^n \, \text{det}( \chi (r)) ) ]^{-1/2}
\exp {\big(- \frac12 \sum_{i=1}^n \frac{\phi_r ^2(r_i)}{ \chi (r_i)} \big)}$,
where $n$ is an arbitrary integer,
and $\chi(r)$ is the covariance (the correlation function)
of the random variable $\phi_r (r_i)$.
When we want to identify a point $r_i$ of the possible path of the wave
according to its amplitude $\phi_r (r_i)$,
the observed amplitude at the point  $r_i$
may be not what we expect since it is a random variable,
thus we do not know if the point  $r_i$ belongs to the path or not.
Thus, we are not able to observe the path of the wave,
nor the distance of the path.
Equivalently,   the  potential $\phi_r$ can be described in $k$-space
as a sum of  infinitely many modes  \cite{BardeenBondKaiserSzalay1986},
$\phi_r (r,t) = 2 \sum_{\bf k}   \phi_r (k,  t )
\cos({\bf k\cdot r} + \theta_k )$,
where each mode $\phi_r (k,  t)$
is a wave traveling  along the  $\bf k$ direction
with a random phase  $\theta_k$ equally distributed on $[0, 2\pi]$.
At a fixed time, $\phi_r (k, t)$ is a random  variable
prescribed by a Gaussian probability distribution,
$P(\phi_r (k)) = \exp {\big(- \frac12  \phi_r ^2(k)/ p_k \big)}$,
where $ p_k$ is the power spectrum.
When we want to identify a wave of  wave number $k$
according to its amplitude $\phi_r(k)$,
the observed amplitude may be not what we expect,
thus we do not know if the observed wave is the one we are seeking.
Moreover,  generally we see  a number of waves with wave numbers close to $k$,
and we can not distinguish them,
due to a limited precision of measurement of $k$.
When we have to pick up one of them arbitrarily,
we will face another problem.
These waves with different random phases may have traveled different distances.
Eventually   we do not know the distance the wave has traveled.
From the above discussion we conclude that,
in both position space and $k$-space for a Gaussian random field,
the traveled paths of waves are wiped out statistically,
and the traveled distance is unobservable,
as is the sound horizon as a traveled distance in the baryon plasma.
(In contrast to the Gaussian random process,
when a piece of stone is dropped in a calm pond,
we are able to observe the path of the wave in perturbed water,
and to measure the traveled distance.
This is,  nevertheless, a deterministic case,
unlike the baryon acoustic waves at the decoupling.)
The situation of BAO at decoupling
is like an instant snapshot of the sea surface full of random waves,
 unlike the calm pond perturbed by a piece of stone.
What we can extract from this photo of ocean surface is
the characteristic wavelengths of ocean waves,
i.e., the power spectrum of  ocean waves,
but not the path of waves from an earlier instant.
Just as Sunyaev and Zel'dovich  \cite{SunyaevZeldovich1970}
 correctly pointed out,
``note that only observations of the small-scale fluctuations of relic radiation
with a periodic  dependence on scale may give information
on the large-scale density perturbations."

The  sound horizon
appears in the  spectrum of BAO that  contains several characteristic peaks
\cite{SunyaevZeldovich1970,PeeblesYu1970,Holtzman1989,HuSugiyama1995},
and the separation between these peaks is approximately
equal to half of the  sound horizon  \eqref{soundhorizon1}
\cite{BashinskyBertschinger2002}.
In this regard,
the sound horizon encoded in the spectrum is an observable,
but not as a distance traveled by BAO random waves.
In our model, the initial power spectrum \eqref{iniPkold}
of the system of galaxies includes one pertinent peak of the BAO spectrum,
so the resulting  correlation function contains part
of the information of BAO spectrum.
But the influence of the peak of BAO spectrum is degenerate with
the other parameters, such as $\lambda_J$, $\gamma$ and the sound speed model.
Our computation tells us that,
from the solution $\xi$,
it is hard to infer the peak of BAO spectrum
to a sufficient accuracy.

(2)   Another  prominent  feature are
the  multiwiggles in the spectrum $P_k(z)$,
which occur  at high $k> k_J$,
as seen in   Fig. \ref{4}, Fig. \ref{5},  and  Fig. \ref{13}.
Firstly $P_k$ has a smooth global maximum plateau,
which appears as the sharp peak in the weighted  $k P_k$
and is  located at $k \simeq  k_J \sim 0.1$Mpc$^{-1}$.
We refer to  it as the main peak.
By Fourier transformation, the main peak gives rise to
the 100 Mpc periodic bumps in $\xi$.
The wiggles  show up on the main slope at high $k> k_J$.
By performing  Fourier transformation,
the  wiggles are not associated with  the periodic bumps of $\xi$,
but rather associated  with the   sub-bumps in \eqref{subbumps},
and are located roughly  at
\bl
k \sim  l k_J, ~~l=2,3,4, ... ,
\el
and their heights are much lower than the main peak.
Both the  main peak and the wiggles
are already observed in galaxy and quasar surveys  \cite{Anderson2014,Ata2018}.
Comparing  the observations,
the overall profile of the  solution $P_k$ agrees with the observed data,
but contains many wiggles at high $k$,
which are expected to  be damped  considerably
when nonlinear terms in  Eq. \eqref{eq2ptcorr34}  are included.

The wiggles are    acoustic oscillations of the fluid with pressure,
occurring  at large $k$ where gravity is subdominant.
This can be also demonstrated analytically from Eq. \eqref{Jseqkm0},
which,  by setting $a=1$ and  dropping  $2H \dot \xi$,
becomes  approximately
the equation of a forced oscillator
\[
\frac{d^2}{dt^2}  P_k  +  c^2_{s0}( k^2 - k_{J}^2 ) P_k
            =4\pi G m .
\]
For      $k >  k_J$, its solution is
\bl
P_k  \sim  b \cos( c_{s0} k \,  t)
        +\frac{4\pi G m }{c^2_{s0}  k^2 },
  \nn
\el
where  $4\pi G m /c^2_{s0}  k^2$
determines the main slope at   $k >  k_J$,
and  $b \cos( c_{s0} k \,  t)$  gives the wiggles,
and the coefficient  $b$ is determined by the initial condition.
The wiggles are oscillating with time- more drastically for higher $k$
-as seen in Figs. \ref{4},  \ref{5},  and  \ref{13}.
The separation between two neighboring wiggles is
$ \Delta k=  \frac{2\pi}{c_{s0} t}$ at a fixed time $t$,
and is narrowing  down during evolution.
Taking into account the  evolution effect,
one has an estimate  $\Delta k \sim 0.1$ Mpc$^{-1}$ at $z=0$,
which agrees with what we see  in the graphs.
Moreover,  the wiggles are developing during evolution
even if the given initial spectrum is smooth without wiggles.
The power spectrum of static solution \cite{Zhang2007,ZhangMiao2009}
does not contain wiggles because the static equation does not contain
the term $\frac{d^2}{dt^2}P_k$.
Thus, given the wiggles at $z=0$,
one can not infer the precise pattern of wiggles at $z=7$,
nor the peak of BAO spectrum,
because other factors,
such as $\gamma$, the details of initial spectrum, etc,
 also affect the outcome  in a complicated way.

\section{ The influences by the expansion,
  the parameters and the initial condition}

We now demonstrate how the solution is influenced by
the expansion, the  sound speed model,
the three parameters ($A,k_J, \gamma$) of the fluid,
and two cosmological parameters ($\Omega_m,h$).

(1)    the influence of  expansion;

For the linear equation in the static case \cite{Zhang2007},
the bump separation  $\Delta r$ is just equal to the Jeans length $\lambda_J$.
But this will be modified in the expanding Universe.
Note that  $\Delta r$  in $\xi$
is contributed to by all the growing $k$-modes
 via  the  Fourier transformation \eqref{Fourierxi}.
In Eq. \eqref{linequavar},
the factor $(\frac{k^2}{k_{J}^2}  - \gamma\,  a^{2\eta -1})$
determines that
the  modes with $ k  <  (\gamma\, a^{2\eta -1})^{1/2} k_{J}$ will grow.
So,  the effective Jeans wave number is
\bl\label{effkj}
k_{J \,   eff} = \frac{ \gamma^{1/2} k_{J}}{(1+z)^{ \eta-1/2}},
\el
which is affected by the expansion,
and  also depends on  $\gamma$  and $\eta$.
Given   $\gamma=1$,  for both models $\eta=1$ and $\eta= 3/5 $,
one has  $k_{J \,  eff} = \frac{ k_{J}}{(1+z)^{ \eta-1/2}} \leqslant  k_J$,
so the growing modes have smaller $k$ than the static case,
and after $k$-integration ,
this leads to a  separation $\Delta r$
which is  larger than $\lambda_J$.

(2) the influence of sound speed model;

\begin{figure}
\centering
\includegraphics[width=0.7\textwidth]{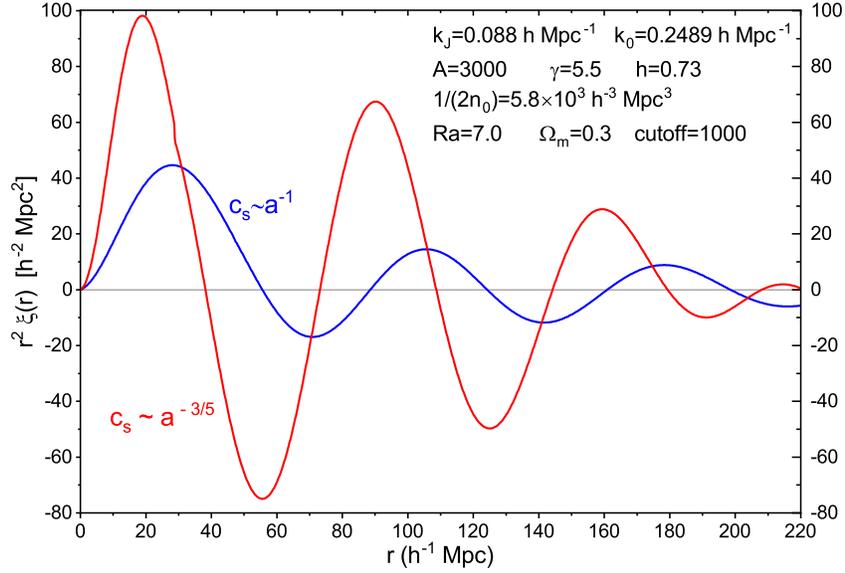}
\caption{The model $\eta= 3/5$  (red line)  yields higher bumps
and shorter separations than the model $\eta= 1$    (blue line).
The initial $k_0=(1+z)^{1/2}k_J$ is  taken.
}
\label{14}
\end{figure}
The sound speed  models  \eqref{csat}  affect the outcome.
By $(c_{s0}/a^{3/5})/(c_{s0}/a)=a^{2/5}<1$,
the   model  $\eta=3/5$,  comparatively,
has a smaller effective sound speed and
thus a shorter Jeans length and shorter bump separations.
Figure \ref{14} shows $r^2 \xi $ in the two models.
In terms  of $P_k$,
the model $\eta= 3/5$  yields a larger  effective Jeans wave number,
  more $k$-modes  will fall into the growing modes
leading to a higher peak of   $P_k$ and higher  bumps.
Besides, the $(k/k_J)^2$ term in Eq.  \eqref{Jseqkm0}
acquires  greater effective coefficients,
so the wiggles of $P_k$ become bigger.
By choosing  the respective  $k_J $ appropriately,
 both models can give  $\Delta r$ of \eqref{sepbump}
that agrees with the observed 100 Mpc feature.

(3) The influence of    $k_J$;

\begin{figure}
\centering
\includegraphics[width=0.7\textwidth]{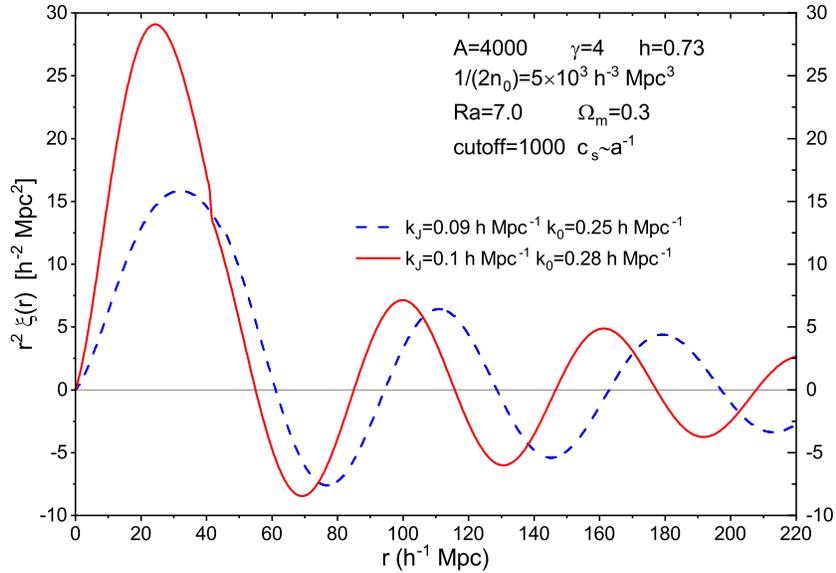}
\caption{   A greater $k_J$ leads to a higher mountain at small scales,
and  shorter bump  separations. }
\label{15}
\end{figure}

By  the definition of $\lambda_J$,
a higher density gives a shorter  $\lambda_J$,
and thus  yields a smaller separation of bumps.
This explains the simulations result \cite{Neyrinck2018,Hernandez-Aguayo2020}
that galaxies residing in dense regions have a shorter bump separation.
A lower  $\lambda_J$ also yields  a mildly higher clustering amplitude,
as seen in Fig.\ref{15}.
This  explains  the simulation result
that galaxies residing in more dense regions give a higher clustering amplitude
\cite{Neyrinck2018,Hernandez-Aguayo2020,SherwinZaldarriaga2012}.

(4)  The influence of   $A$;

The source magnitude $A$ is proportional to $m$.
Our computation shows that
a larger $m$ gives a higher main mountain at $r \lesssim 50$Mpc,
but  does not affect the periodic  bumps at large distance.
In particular,  the solution for $A=0$ still contains bumps at large distance.
To explain this novel phenomenon,
we  decompose  the solution $\xi$ into two parts
\bl \label{xi1xi2}
\xi  = \xi_1  + \xi_2,
      ~~\text{also}~~
P_k= P_{k\, 1} + P_{k\, 2},
\el
where $\xi_1$ is the inhomogeneous solution with $A\ne 0$
and the zero initial condition   ($P_{k\, ini}= r_a=0$),
and $\xi_2$ is the homogeneous solution with $A=0$
and the nonzero initial condition.

\begin{figure}
\centering
\includegraphics[width=0.7 \textwidth]{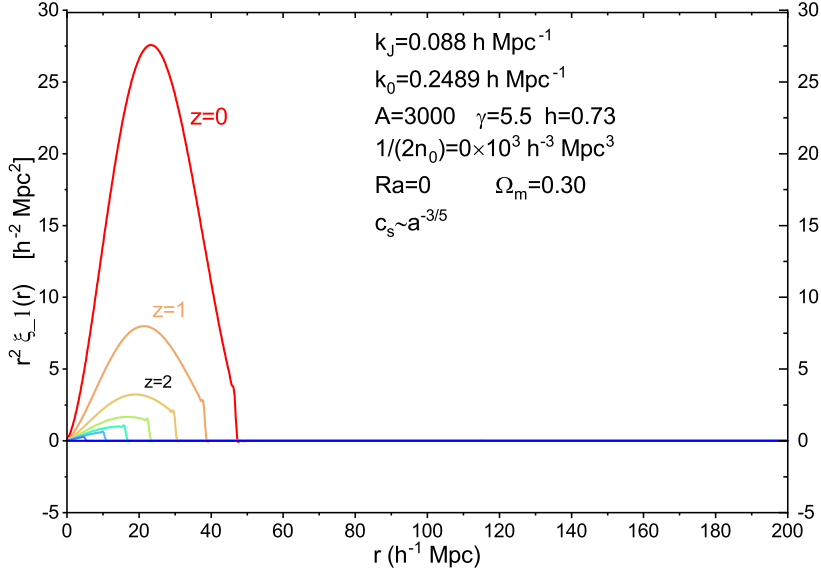}
\caption{The inhomogeneous  $r^2 \xi_1$
generates the main mountain at small $r$ which  is growing.
There  are no bumps at large $r$ beyond the mountain.
}
\label{16}
\end{figure}

\begin{figure}
\centering
\includegraphics[width=0.7 \textwidth]{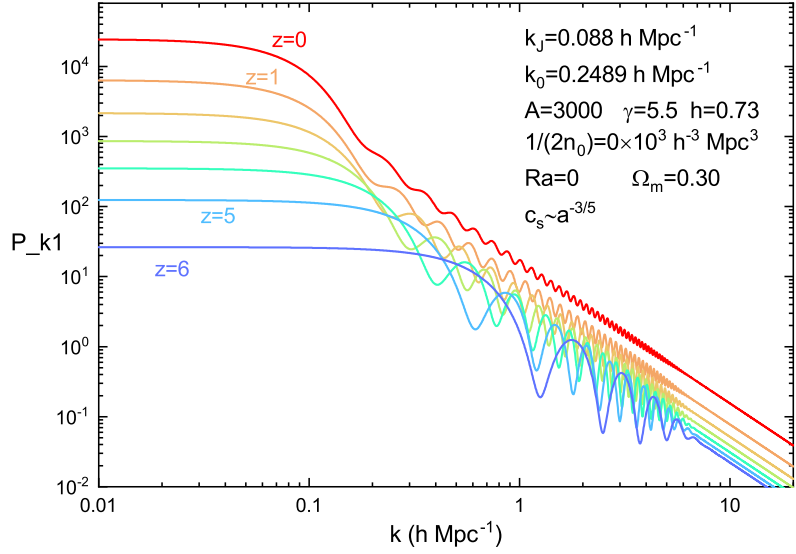}
\caption{ Same as Fig. \ref{16}.
The   inhomogeneous $P_{k\, 1}$ has a flat main peak at small $k$
which is shrinking to small $k$ during evolution.
Multiple wiggles are developing on the main peak at large $k$.
}
\label{17}
\end{figure}

$\xi_1$ and $P_{k\, 1}$  reflect the influence of  $A$,
and are shown in    Fig.\ref{16} and Fig.\ref{17}.
$\xi_1$ gives the growing main mountain,  but  contains no  bumps.
At any instance of time, $\xi_1$ is vanishing beyond the main mountain,
so it describes the local clustering around the galaxy  with mass $m$.
The main mountain is growing radially at the sound speed.
 $P_{k\, 1}$ is  flat and smooth without a sharp edge at small $k$,
and this explains why here are  no bumps  in $\xi_1$.
Moreover, $P_{k\, 1}$  is developing multiple wiggles at large $k$ during evolution,
even though  the initial spectrum is zero.
This tells  us  that  the wiggles do not give rise to the periodic bumps in $\xi$.

\begin{figure}
\centering
\includegraphics[width=0.7\textwidth]{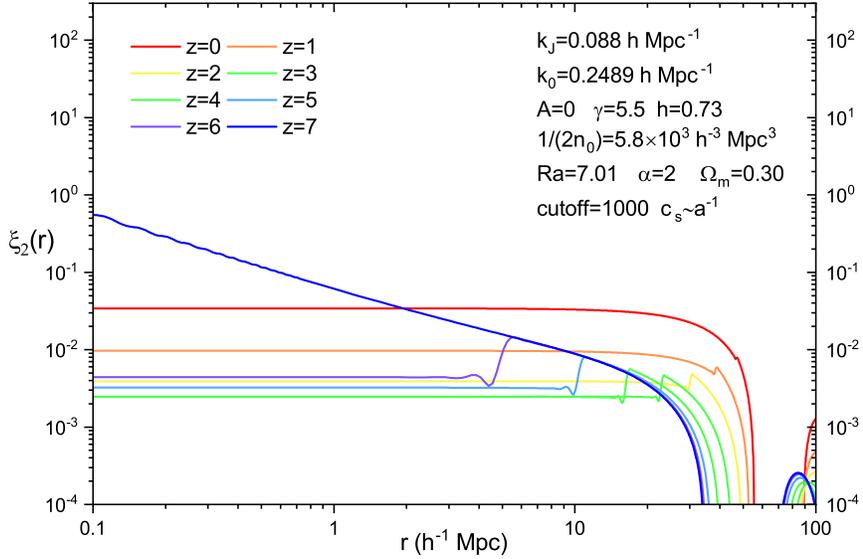}
\caption{The  homogeneous solution $\xi_2(r,z)$
forms a plateau on small scales ($r<40\,\mathrm{Mpc}$)
 at late times ($z\sim 0$).
}
\label{18}
\end{figure}

\begin{figure}
\centering
\includegraphics[width=0.7\textwidth]{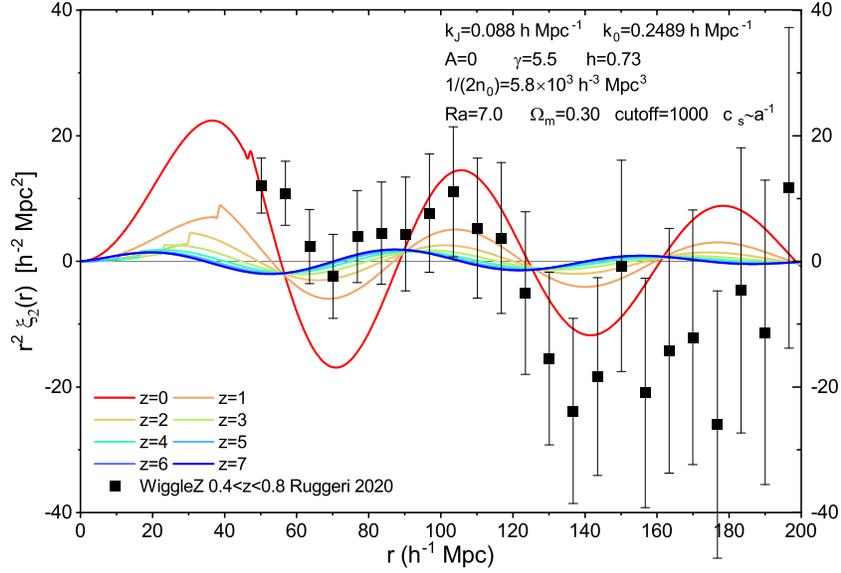}
\caption{ Same as Fig.\ref{18}.
The  weighted  $r^2 \xi_2(r,z)$ shows the periodic bumps at large  scales.
}
\label{19}
\end{figure}

\begin{figure}
\centering
\includegraphics[width=0.7\textwidth]{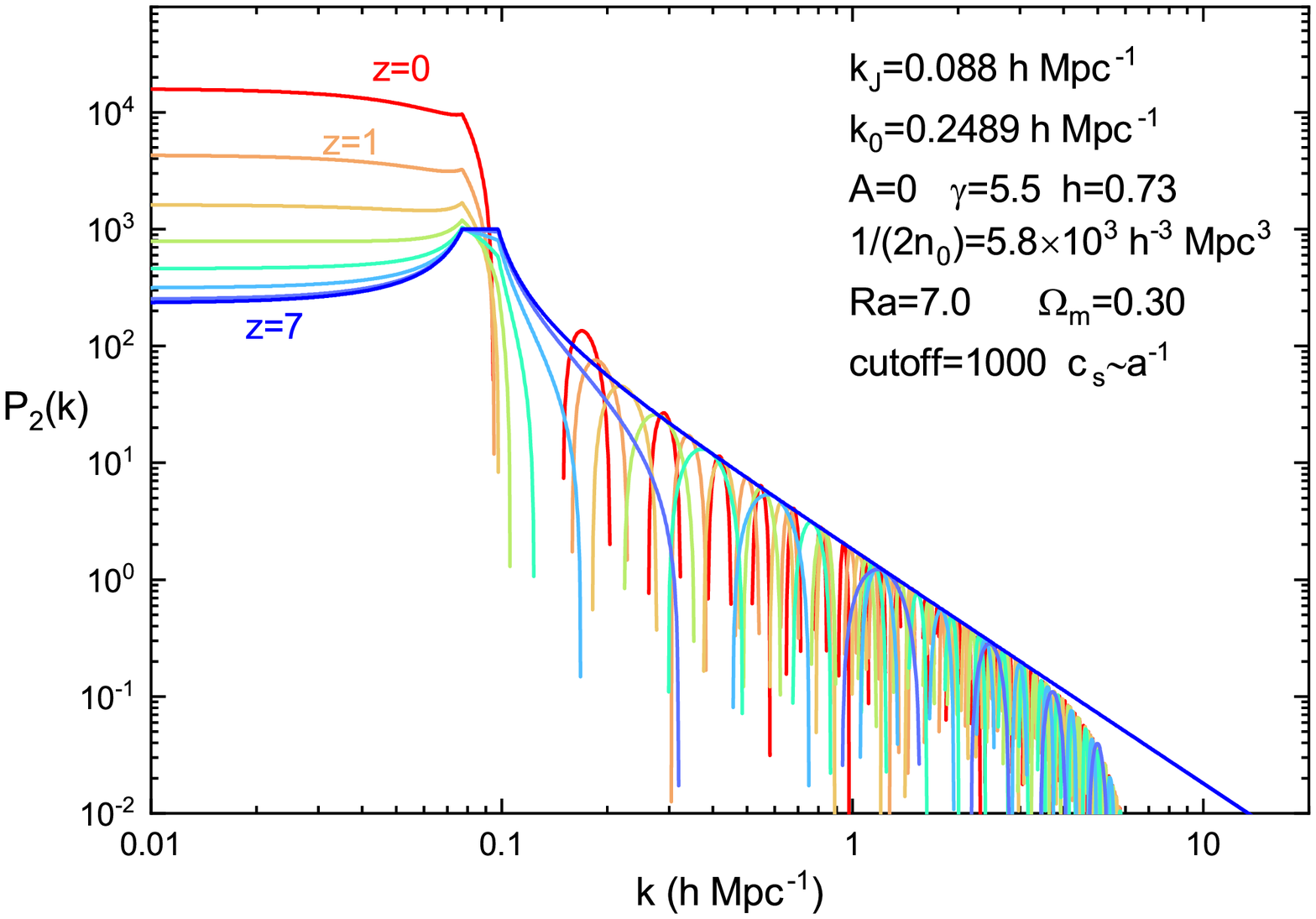}
\caption{  Same as Fig.\ref{18}.
The homogeneous power spectrum $P_{k\, 2}(z)$
has a sharp main peak at $k\lesssim k_J$, but little power at large $k$.
}
\label{20}
\end{figure}

The homogeneous part $\xi_2$ and $P_{k\, 2}$
 reflect the influence of the nonzero initial condition and
are shown in  Fig.\ref{18},  Fig.\ref{19}, and Fig.\ref{20}.
$\xi_2$  contains the periodic bumps at large scales, but
forms a flat plateau on small scales $r< 40$Mpc at late times.
 $P_{k\, 2}$ has the main peak with a sharp edge at $k\sim k_J$
which gives rise to the periodic bumps of $\xi$,
but has little  power at large $k$,
corresponding to the flat plateau in $\xi_2$.
This reconfirms that the bumps in $\xi$ are associated with
the main peak of $P_k$, not with the wiggles in $P_k$.
 Moreover,
the evolution of $\xi_2$ in Fig.\ref{19} also shows
that the bumps are distributed over the whole $r$ axis,
and the bumps are getting higher, the troughs are getting deeper,
and the bump separation is getting larger during evolution.
This behavior of $\xi_2$ for the large scale structure
 differs from that of $\xi_1$ for the local clustering.
It should be mentioned that the main mountain of $\xi$
is a superposition of $\xi_1$ and the first bump  of $\xi_2$.

From  the decomposition in the above,
we conclude that
the main mountain of $\xi$  at  $r \lesssim 50$ Mpc is due to the source $A$,
while  the periodic bumps at large distance
are seeded by the nonzero initial condition \eqref{iniPkold}.
Thereby,   at the linear level the small scale clustering and
the large scale structure are
separated into are two different problems.
In Appendix B we also express the decomposed solutions
in terms of the Green's function,
and demonstrate the wave nature of correlation function.

(5)  The influence of  density ratio  $\gamma$;

\begin{figure}
\centering
\includegraphics[width=0.7\textwidth]{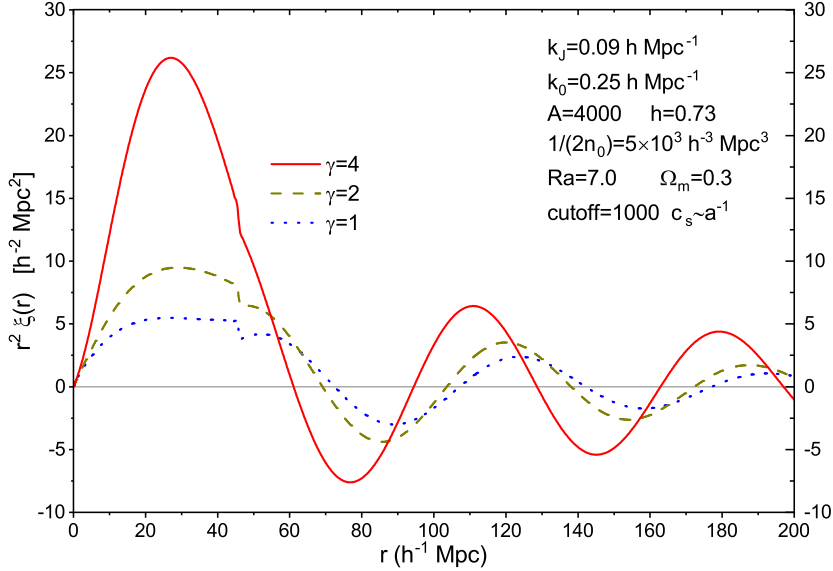}\\
\caption{
A greater  $\gamma$ gives  higher bumps and deeper troughs
and shifts the bumps to a smaller distance.
}
\label{21}
\end{figure}

The  ratio $\gamma$ defined by Eq. \eqref{defgamma}
is the fluid density over the cosmic background matter density,
and can be regarded as the region overdensity of survey
over the  background density.
A larger $\gamma$ yields higher bumps and deeper troughs,
and simultaneously shifts the bump locations of $\xi$ to small distance,
as   shown in  Fig. \ref{21}.

(6) The dependence on $\Omega_m$;

\begin{figure}
\centering
\includegraphics[width=0.7\textwidth]{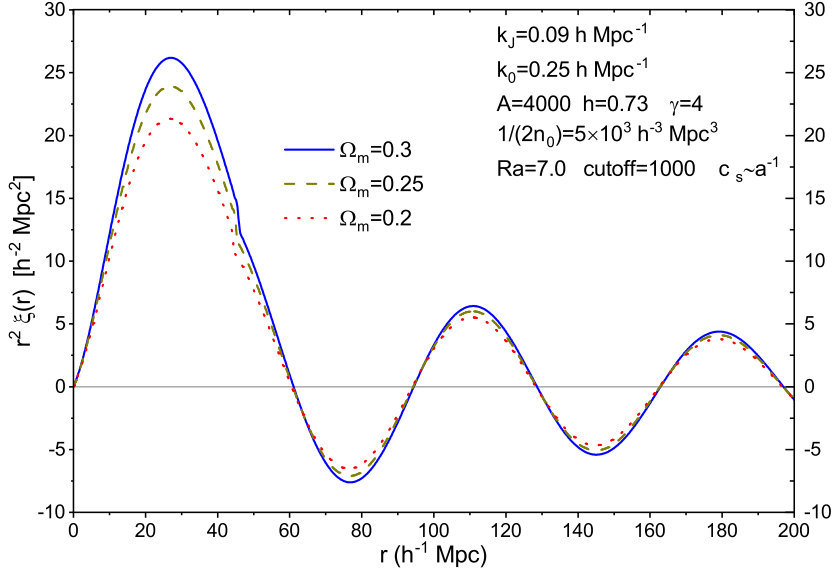}\\
\caption{A high $\Omega_m$ gives slightly higher bumps at $z=0$,
but makes no change in the separation between bumps.
}
\label{22}
\end{figure}

The matter fraction $\Omega_m$ of the cosmic background
also affect slightly  the correlation and clustering.
A high $\Omega_m$ enhances  slightly the height of the bumps,
but does not change the bumps separation,
as seen  in    Fig.\ref{22},
where  $k_J$ and $\gamma$ are fixed.
$\Omega_m$ actually occurs in the definition of $k_J$ in Eq. \eqref{kJdef}.
If we would allow $k_J$ to vary  with   $\Omega_m$,
then a larger  $\Omega_m$ will correspond to a larger $k_J$
and will lead to a shorter bump separation.

(7) The dependence on   $h$;

The Hubble parameter $h$ occurs in the  source amplitude  $4\pi G m /H_0^2$
of Eq. \eqref{linequavar}
and in  the  definition of $k_J$ in \eqref{kJdef}.
So a small $h$ amounts to a greater $m$ and a greater $k_J$.

We now demonstrate the influence of
the initial condition  \eqref{iniPkold}  and \eqref{raterv}.
Beside the parameters,
the initial power spectrum is another important factor
 that affects the solution  of  Eq.  \eqref{linequavar}.
To get a smooth solution,
the initial condition should be in a range in accordance with the given parameters.

(1) the influence of initial Jeans wave  number $k_0$;

\begin{figure}
\centering
\includegraphics[width=0.6 \textwidth]{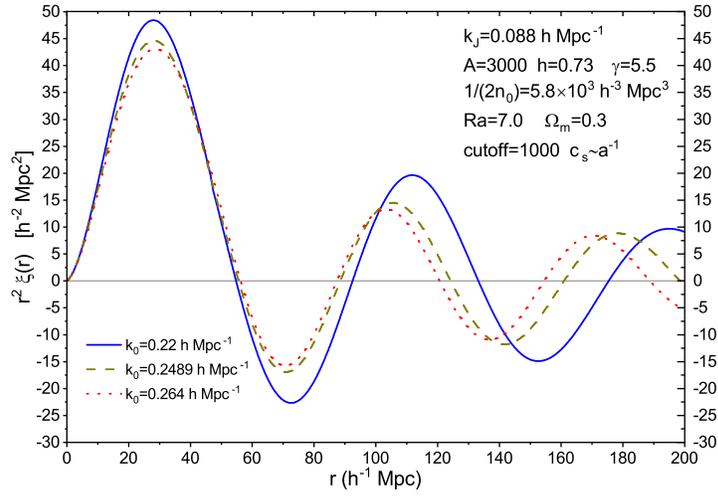}
\caption{ A greater $k_0$ leads to  lower bumps and shorter bump separations.}
\label{23}
\end{figure}

Our computation shows  that
it is necessary that the initial spectrum possess certain sharp peak or wedge
which is located at  $k_0$,
for the periodic bumps to develop in $\xi$.
As mentioned in Sect 4,
the initial peak position $k_0$ can be viewed as an imprint
of the peak of the BAO spectrum,
and its value is related to $k_J$
through the relation \eqref{k0K0}  by default.
If  $k_0$  varies slightly from the relation,
then a greater $k_0$ leads to  lower bumps and shorter bump separations
in $\xi(r)$, as shown in  Fig. \ref{23}.
We have seen that
the influence of $k_0$ is degenerate with the parameters $\gamma$ and $\eta$
to various extents,  on the outcome $\xi$ and $P_k$.
As our computation shows,
$k_0$ should not  be allowed to deviate too much,
otherwise the evolution may  be not sufficiently smooth.

(2)  the influence of initial amplitude $1/2n_0$;

\begin{figure}
\centering
\includegraphics[width=0.6 \textwidth]{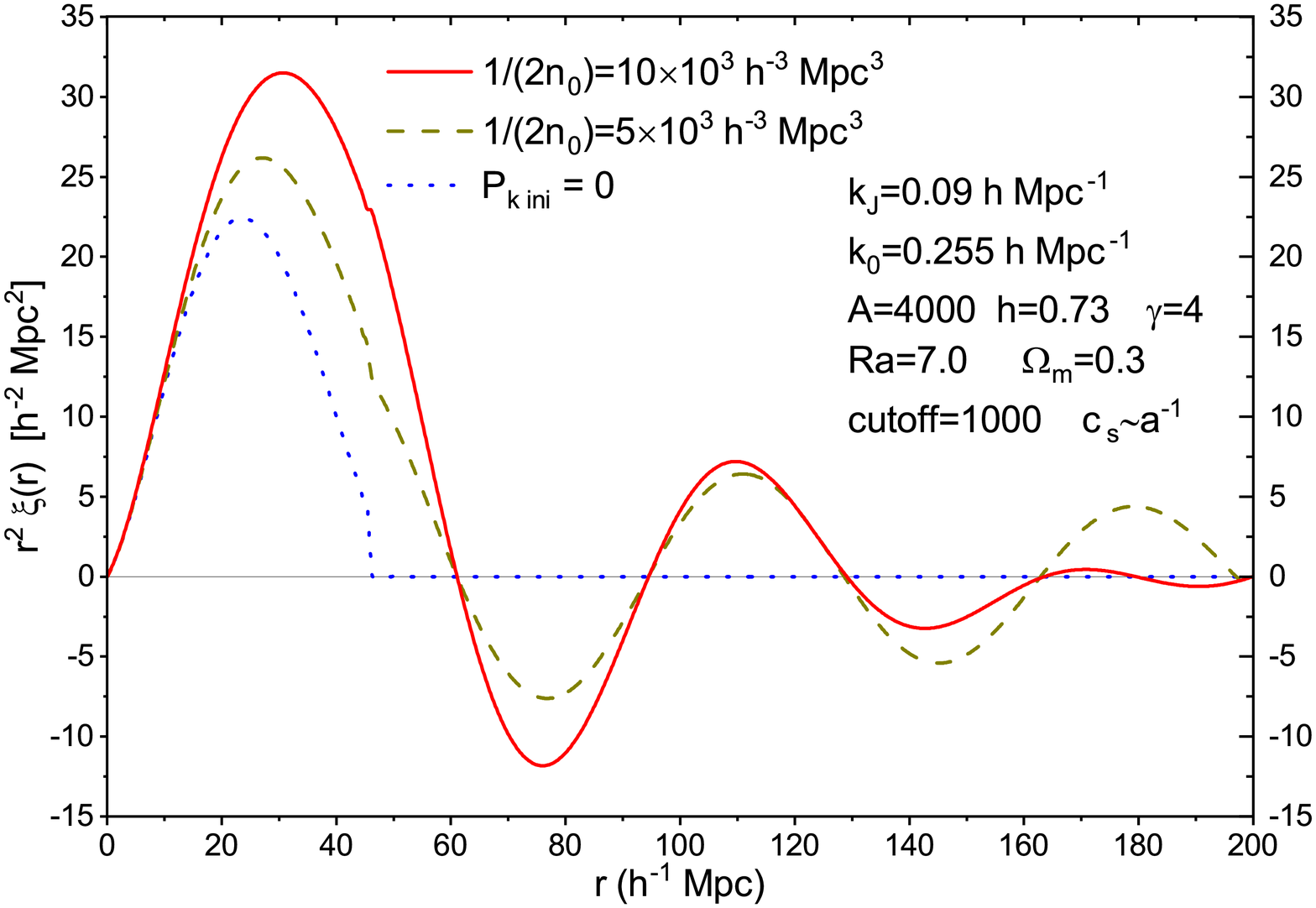}
\caption{ The influence of  $1/2n_0$ upon  $r^2 \xi$.
A higher  $1/(2n_0)$  yields a higher amplitude at $r\lesssim 50$ Mpc.
In particular,
when the initial power spectrum is vanishing,  $P_{k\, ini}=0$,
the solution is contributed only by the inhomogeneous  $\xi_1$.
There are no oscillatory bumps  in $\xi$ at large $r$,
but the main mountain still exists at small scales.
}
\label{24}
\end{figure}

$1/2n_0$ is used to represent
the amplitude of the initial power spectrum in \eqref{iniPkold}.
A higher  $1/(2n_0)$  yields a higher amplitude at $r\lesssim 50$Mpc
and a deeper first trough in $\xi$,
but leads to a complicated pattern for the subsequent bumps and troughs,
as seen  in Fig. \ref{24}.
The main peak of $P_k$ is also slightly enhanced
but the wedge also shifts to large $k$,
and there is little change at large $k\gtrsim   0.2$ Mpc$^{-1}$.
In the limiting case of a vanishing initial power spectrum, $P_{k\, ini}=0$,
the solution is the inhomogeneous  $\xi_1$
that  has  no bumps and  has been  analyzed  below Eq. \eqref{xi1xi2}
and in  Appendix B.

(3)  the influence of initial rate  $r_a$;

The default  initial rate is $r_a=7.0$ by \eqref{raterv}.
We find that the evolution pace and outcome
are not sensitive to the initial rate $r_a$
within two orders of magnitude.
This is because  in  Eq. \eqref{linequavar}
the coefficient of $\frac{\partial}{\partial a} P_k$
is given by $\sim \frac{3}{2a}$,
which is about  three orders lower than the coefficient of $P_k$ term at $z=7$,
so that a small variation of  $r_a$ gives no substantial change to the solution.
This also means that the impact of expansion
is subdominant to those of  pressure and gravity
for the system of galaxies
during the epoch from $z=7$ to $z=0$.

\section{Conclusion and Discussion}

Based on  the hydrodynamical equations of a Newtonian self-gravity fluid,
we   derived the nonlinear equation of the  correlation function of
density perturbation in a flat expanding universe.
This  extends our previous work in the static universe
\cite{Zhang2007,ZhangMiao2009}.
Our nonlinear equation
is compared with Davis and Peebles' equation \cite{DaviesPeebles1977}
in the following.

Like  Davis and Peebles,  we use  a self-gravity fluid model
to describe the system of galaxies in the expanding Universe.
Starting from  statistical distribution,
Davis and Peebles start with the Liouville's equation and derive
a series of BBGKY equations of two-point  correlation functions with cutoff.
This is a standard method for many-body systems.
We describe the system of galaxies by the density $\psi$
which is a stochastic field,
apply  functional differentiation
on the ensemble average of the density field equation \eqref{eqevpsiJ},
perform expansion in terms of density perturbation,
and obtain Eq. \eqref{eq2ptcorr} of two-point  correlation function.
Our method is commonly used in field theory,
and the derivation involves less algebraic calculations.
Our Eq.  \eqref{eq2ptcorr} is analogous to  Davis and Peebles'  Eq. (47),
except there is a factor of two  for the gravity term \cite{DaviesPeebles1977}.
From   the outcome, the BBGKY series is  effectively  equivalent to
the expansion of density perturbations that we have used.
We have assumed the generating functional \eqref{ZJdef} for the density field,
which is also a prescription of the statistic of the field.
The resulting evolution of correlation function is smooth,
and the large-scale structure keeps a similar pattern
during evolution  $z=(7\sim 0)$.
In this sense, the system is  in an asymptotically relaxed state
\cite{Saslaw1985}.

There are  several   differences.
To deal with the velocity terms that occur in  Eq.  \eqref{eq2ptcorr},
we have used the Zel'dovich approximation
to replace the velocity by an integration of the density field,
and arrived at Eq.  \eqref{eq2ptcorr1}
which contains three-point and four-point  correlation functions.
This hierarchy is generally  anticipated
for a many-body system  with interaction and  also for a nonlinear field.
To deal with  three-point and four-point   correlation functions,
we adopt the Kirkwood-Groth-Peebles ansatz and the Fry-Peebles ansatz,
and obtain the main Eq.  \eqref{eq2ptcorr34},
which is a closed equation for the two-point correlation function
as a single unknown function.
Davis and Peebles  \cite{DaviesPeebles1977},
without using the Zel'dovich approximation,
derived the differential equations of the velocity and of velocity dispersions,
and arrived at a set of five equations  (71a), (71b), (72), (76), and (79)
for five unknowns.
For practical application,
one  needs to to specify an initial condition
which should be  consistent  for the five unknowns.
This would not be easy
since one generally lacks sufficient information of these quantities at early epoch.
Moreover, the pressure term $c_s^2 \nabla^2 \xi$
and the  $\delta^{(3)}$  source term
were ignored in Davis and Peebles' final equations,
so that the acoustic properties (the periodic bumps and wiggles)
and the local clustering   at small scales (the main mountain)
would not appear.

Our Eq.  \eqref{eq2ptcorr34}  is  nonlinear, also integro-differential.
In this paper we only solve its linear version,
ie, the linear equation \eqref{linapprg}, apply it to the system of galaxies.
For specific computation,
we adopt two models of  $c_s$ in \eqref{csat}.
The initial power spectrum \eqref{iniPkold} at $z=7$
is taken from that of the static  solution,
and  inherits a portion of the imprints of the BAO spectrum
that has survived the Silk damping around the decoupling.
The initial  rate $r_a$ is specified by \eqref{raterv} based on the pair conservation.

The linear solution  $\xi$ contains a power-law main mountain at small scales
and the periodic bumps at large scales.
These were previously predicted in the static solution,
and  confirmed by the observations.
Taking  advantage of the  linearity,
we also decompose the solution into homogeneous and inhomogeneous solutions,
$\xi=\xi_1+\xi_2$,
and analyze these solutions in terms of  Green's function.
$\xi_1$  is proportional to $m$,
and generates the growing main mountain,
and $\xi_2$  gives rise to the growing periodic bumps which are unaffected by $m$.
Thereby,  the local clustering and the large-scale structure are
naturally separated as two problems.
The bump separation $\Delta r \sim 100$ Mpc
is largely determined by the Jeans length $\lambda_J$
but also modified by $\gamma$, the sound speed model,
and the initial condition.

The corresponding power spectrum $P_k$  contains
a main peak which is associated with the periodic bumps,
 as Fourier transformation shows.
$P_k$ also contains  the multiwiggles
which are caused by the acoustic oscillations of the system of galaxies.
The wiggles are absent in the static solution,
and, nevertheless,
are developing during evolution
even if the initial spectrum has no wiggles.
The wiggles do not generate the 100 Mpc periodic bumps,
but rather generate the tiny sub-bumps in $\xi$ which are barely visible.
In this perspective,
$P_k$ gives information of acoustic oscillations more  than $\xi$ does.
Given the pattern of  wiggles at present stage,
we are not able to accurately infer the wiggle pattern at $z=7$,
because other factors also affect the outcome in a complicated way.

The sound horizon,
when interpreted as a distance that baryon acoustic waves traveled,
is not directly observable from the plasma.
This is because the waves form a Gaussian random field,
and the paths of waves are wiped out statistically.
Indeed, the predicted value of the  sound horizon as a distance ruler
is higher than the observed 100 Mpc feature,
and can not give a simple account of the negative trough at ($130 \sim 160$)Mpc,
nor  the second bump at $\sim 200$ Mpc.
Therefore,
 the conventional picture of the comoving imprint of sound horizon has a difficulty.
The sound horizon actually occurs in the phase of BAO modes
-one half of sound horizon  is approximately equal to
the separation between the characteristic peaks of the  BAO spectrum.
In our present model,
the imprint of BAO is transferred to the initial power spectrum
 of system of galaxies, say,  at $z=7$.
Subsequently the pertinent quantity is the Jeans wavelength
which departures from comoving.
The separation between the observed 100 Mpc bumps
is attributed to  the Jeans length,
and is also modified by the parameters  $\gamma$ and $c_s$,
so it is hard to accurately infer
the BAO spectrum from the outcome  $\xi$ and $P_k$ at $z =0$.
When future surveys provide more precise measurement
 of the second bump at $\sim 200$Mpc,
or even observe the third bump at  $\sim 300$Mpc,
it would be possible to infer  information of the  BAO spectrum.
It should be noticed that
the measurement of the Hubble constant using the  CMB
 without using the sound horizon \cite{Spergel2007,Dunkey2008}
has yielded the result $h\sim 0.72$ which is consistent with the local measurements
\cite{Reid2019,Pesce2020,Wong2019},
while those, using the sound horizon,
lead to an underestimate  $h\sim 0.67 $ \cite{Ade2015,Aghanim2018},
in sharp contrast to the local measurements.
Therefore,
the conventional use of the sound horizon as a ruler
for cosmological distances  should be reexamined for BAO and for CMB as well.
Given the Hubble tension between the local measurements and
the sound-horizon-based CMB+BAO,
this scrutiny is  necessary.

Another important lesson from the  linear solution is that
the Jeans length is the correlation scale of the system of  galaxies,
and is distinguished from the mass scale of a galaxy or cluster.
Since the pioneering studies of BAO around the decoupling
\cite{SunyaevZeldovich1970,PeeblesYu1970,Silk1968,Field1971,Weinberg1971},
the Jeans length \cite{SunyaevZeldovich1970}
and the characteristic peaks in the spectrum \cite{PeeblesYu1970},
were often thought to be associated with a Jeans mass
 enclosed by the associated volume,
such as $10^{17}M_\odot$ or $10^{5}M_\odot$,
in hope for an interpretation of  the origin of galaxies and clusters.
This is a longstanding problem in cosmology and galaxy formation.
According to our linear solution,
the imprint of BAO at the decoupling
evolves into the present Jeans length $\lambda_J$
which is roughly equal to the bump separation  $\Delta r$,
and the mass enclosed in a sphere of radius $\frac12 \Delta r$  is
given by
$\frac{\pi}{6}\rho_c \Omega_m (   \Delta r)^3 \sim   10^{17}M_{\bigodot}$,
just comparable to the Jeans mass  predicted
by  Refs.\cite{SunyaevZeldovich1970,PeeblesYu1970,Weinberg1971,Silk1968,Field1971}.
Obviously, this Jeans mass does not correspond to the mass scale of a galaxy,
 nor of a cluster.
The decomposition \eqref{xi1xi2} of the  solution reveals
that the mass scale of a galaxy is described by $m$
which is responsible for  the local clustering,
and that $\lambda_J$ occurs not as a (virilized) galactic object,
but as the correlation length of galaxies,
which is responsible for the large scale structure.
The two  parameters
$\lambda_J$ and $m$ reflect two different aspects
of the system of galaxies in the Universe.
Thereby, one realizes that,
in the fitting formula of observed correlation function $\xi =(r/r_0)^{-1.7}$,
the constant $r_0$ is really not a correlation length,
but a phenomenological parameter
which reflects the effects of $m$ and nonlinearity at small scales.

Obviously the linear solution has  shortcomings on small scales.
The slope of main mountain of $\xi$  is too flat,
leading to an overestimated mass $m$ for a galaxy.
The predicted wiggles in  $P_k$ at high $k$  are undamped.
These are due to the absence of nonlinear terms
from Eq.  \eqref{linapprgtcs0},
which are dominant at small scales.
In future,  we shall study the nonlinear effects.

\section*{Acknowledgements}

Y. Zhang is supported by National Natural Science Foundation of China,
Grants No. 11675165,  No. 11633001, and No.  11961131007,
and in part by National Key RD Program of China (2021YFC2203100).

\appendix

\numberwithin{equation}{section}

\section{Derivation of the nonlinear equation of correlation function}

We first list the field equation of mass density
which is the basis for the equation of correlation of density perturbations.
The system of galaxies is   described
by a Newtonian self-gravity fluid,
whose  hydrodynamical equations consist of the following \cite{LandauLifshitz,Peebles1980}
\bl
\frac{\partial \rho}{\partial t} +\nabla_{\bf r} \cdot (\rho {\bf V})
   & =0, \label{eqrho} \\
\frac{\partial \bf V}{\partial t}    +({\bf V}\cdot \nabla_{\bf r}){\bf V}
           &  =-\frac{1}{\rho}\nabla_{\bf r} p - \nabla_{\bf r} \Phi,  \label{eqv}
 \\
\nabla^2_{\bf r} \Phi &   = 4\pi G \rho  -\Lambda,    \label{poisson}
\el
where $\bf r$ is the proper distance from some chosen origin,
$\bf  V$ is the proper velocity,  and $\Phi$ is the gravitational potential.
The pressure $p$ is comparatively small and its contribution
to the potential is neglected,
but  the  pressure gradient is included in the Euler equation
to show the acoustic behavior of the fluid.
These equations are  valid  for a static universe
within the framework of Newtonian gravity.
To pass to the expanding Universe,
we write $\bf r$ as  ${\bf{r}} = a(t)\bf x $
where  $ \bf x $  is the  comoving coordinate.
(We choose $a=1/(1+z)$, so that ${\bf r}= {\bf x}$ at $z=0$.)
Then one has the  following
\be \label{transf}
\nabla_{\bf r}=\frac{1}{a}\nabla_{\bf x}, \ \ \ \
\left(\frac{\partial}{\partial t}\right)_{\bf r}
=\left(\frac{\partial}{\partial t}\right)_{\bf x} - \bf{ V_0} \cdot \nabla_{\bf r},
\nn
\ee
\[
{\bf V} =\dot { \bf r }=   {\bf V_0} + {\bf v} ,
\]
where ${\bf  v }\equiv a \dot {\bf x}$ is the peculiar velocity field,
and  ${\bf V_0}    =H \bf r$ is the Hubble flow velocity.
We introduce a new potential
\be\label{phiPhi}
\phi  \equiv  \Phi +\frac{1}{2}a\ddot a  x^2  .
\ee
Then, Eqs.\eqref{eqrho},  \eqref{eqv}, and  \eqref{poisson} become \cite{Peebles1980}
\bl
& \frac{\partial \rho}{\partial t}+3\rho H
      +\frac{1}{a} \nabla_{\bf x}   \,   \rho \cdot {\bf   v}
      +\frac{1}{a} \rho(\nabla_{\bf x}  \cdot{\bf  v})=0 \, , \label{continuityexpan}
      \\
&   \frac{\partial    {\bf v} }{\partial t} + H    {\bf v}
          + \frac{1}{a}(  {\bf v}  \cdot \nabla_{\bf x} )   {\bf v}
  =   - \frac{1}{a \rho}\nabla_{\bf x}  p - \frac{1}{a}\nabla_{\bf x} \phi \, , \label{Eulerexpan}
         \\
&   \nabla^2_{\bf x}   \phi =   4 \pi G a^2 \big(\rho -\rho_0(t) \big) \, , \label{Poissonphi}
\el
where $\rho_0(t)$ is the mean  mass density of the fluid,
and $\phi$ is contributed only by the matter density fluctuation.
We denote $\nabla \equiv\nabla_{\bf x}$.
From \eqref{continuityexpan}, \eqref{Eulerexpan}, and  \eqref{Poissonphi},
one obtains the nonlinear field equation  of mass density
\be\label{equrho}
  \ddot \rho  +8H  \dot \rho   +  (15 H^2 +3\dot H) \rho
=  \frac{1}{a^2  } \nabla^2   p
   +\frac{1}{a^2}\nabla  \cdot ( \rho \nabla \phi )
   +\frac{1}{a^2} \frac{\partial^2 }{\partial x^\beta \partial x^\alpha}
               ({\rho  v^\alpha} v^\beta) ,
\ee
which  holds a flat   expanding universe.
Notice that  Eq. \eqref{equrho} has been derived
without using the Friedmann equations explicitly,
and the fluid mass density  $\rho$ in Eq.\eqref{equrho}
can be generally higher than
the background  density of the expanding Universe.
Introducing  the dimensionless mass density $\psi$ as the following
\bl
\rho(t, {\bf x}) \equiv \rho_0(t) \psi(t,{\bf x}),
\el
then  Eq. \eqref{equrho} is written as  Eq. \eqref{eqevpsiJ}.
Defining   the density contrast $\delta$ as
\be \label{defdelta}
\delta({\bf x}, t) = \psi({\bf x}, t) - 1,
\ee
Eq.\eqref{equrho} can be also expressed
\be\label{nonlindelta}
\frac{\partial^2 \delta}{\partial t^2} +2H \frac{\partial \delta}{\partial t}
=  \frac{1}{a^2 \rho_0 } \nabla^2   p
  +4\pi G   \rho_0 ( \delta ^2  +   \delta )
  +\frac{1}{a^2} \nabla\delta \cdot \nabla \phi
   +\frac{1}{a^2} \frac{\partial^2 }{\partial x^\beta \partial x^\alpha}
               \big( (1+\delta)   v^\alpha  v^\beta \big)  ,
\ee
which is  Eq. (9.19) in Ref. \cite{Peebles1980}.
When the three nonlinear terms    are neglected,
\eqref{nonlindelta} reduces to
\be
\frac{\partial^2 \delta}{\partial t^2} +2H \frac{\partial \delta}{\partial t}
  -  \frac{1}{a^2 \rho_0 } \nabla^2   p
 - 4\pi G   \rho_0   \delta =0 ,
\ee
which is the Jeans linear equation  in the expanding Universe.

The field equation of  the two-point  correlation function in the expanding Universe
can be derived by the  functional derivative method,
in a similar procedure to
the static case \cite{Zhang2007,ZhangMiao2009}.
The method  is   commonly used
to get the equation of the two-point  correlation function  in field theory,
such as particle physics and condensed matter
\cite{BinneyDowrickFisherNewman1992,Goldenfeld1992,Zustin1996}.
The system of galaxies in the expanding Universe is
 not too far from equilibrium.
The cosmic expansion time scale $t_e = \frac{1}{H_0}= \sqrt{3/8\pi G \rho_0}$,
the dynamic time for galaxies moving in the background $t_d = \sqrt{3/16\pi G \rho_0}$
 \cite{BinneyTremaine1987},
and the two time scales are  of the same order of magnitude.
So the system of galaxies is said
to be in an asymptotically relaxed state \cite{Saslaw1985}.
In statistical mechanics, given the Hamiltonian  of
    a self-gravitating many-body system,
   the grand partition function is commonly constructed
   with the temperature $T$ \cite{Saslaw1985}.
   By the Hubbard-Stratonovich transformation  \cite{Zustin1996},
   the grand partition function of the discrete many-body system
   is cast into a path-integral generating functional,
   either for the gravitational potential field \cite{deVega1996a},
   or  for the density field $\psi$  \cite{Zhang2007,ZhangMiao2009}.
The generating functional
facilitates the derivation of correlation functions of the field.
We assume that,  in the expanding Universe,
the generating functional of the density field $\psi$
has the following form
\bl
Z[J]    &  = \int {\cal D} \psi
    \exp \Big[ - \beta \int d^3 x
    \big(   {\cal L}(\psi )   -   a^3  J \psi \big)    \Big] ,
    \label{ZJdef}
\el
and the ensemble average of the  field $\psi$ is given by
\bl
\langle \psi \rangle_J
=  \frac{1}{Z[J]} \int {\cal D} \psi \,  \psi   \exp \Big[ - \beta \int d^3 x
    \big( {\cal L}(\psi ) -  a^3 J \psi \big) \Big] ,  \label{avergdef}
\el
where ${\cal L}$ is the effective Lagrangian density
whose  variation  with respect to    $\psi$ leads to Eq. \eqref{eqevpsiJ},
 $J$ is the external source introduced
as an apparatus for functional differentiation,
and $\beta \equiv  1/4\pi G m$ with $m$ being the particle mass.
Although $\beta$ formally plays a role of an  "effective" temperature
in $Z[J]$ of \eqref{ZJdef},
actually it is not the temperature $T$,
the latter has been absorbed into the sound speed $c_s$.
The notation $\beta$ here is different from
that in Refs.\cite{Zhang2007,ZhangMiao2009},
and $c_s^2$ is put into ${\cal L}$ here.
In the static case the explicit expression of ${\cal L}$ is  known
 \cite{Zustin1996,deVega1996a,Zhang2007,ZhangMiao2009}.
In the expansion  case the linear part of ${\cal L}$ is given by
\[
{\cal L }_{lin}  =  a^3
   \Big( \frac12 a^{-1} (\dot \psi)^2 - \frac12 a^{-1}(\frac{c_s}{a}\nabla \psi)^2
    +  4\pi G\rho_0 a^{-1}(\frac{1}{2}\psi-1)\psi \Big),
\]
which corresponds to the part of  Eq.  \eqref{eqevpsiJ}
without the potential and velocity terms.
The nonlinear part of ${\cal L}$ gives rise  to
the potential and velocity terms in \eqref{eqevpsiJ}
and will be more involved,
and we do not need
its explicit expression  in this paper.
Generally speaking,  knowing the  exact expression
of ${\cal L}$ in terms of $\psi$
would amount to knowing the exact nonlinear equation
and the non-Gaussian statistic of the field $\psi$.
In particular,
the non-Gaussian statistic
can be represented by various correlation functions of $\psi$
to a sufficient order.
Therefore,  we want to know the equations of these correlation functions
which contain both statistical and dynamical information
of the field.
The prescription \eqref{ZJdef} with \eqref{avergdef}
of the generating functional is sufficient for our purpose to
derive the equations of various correlation functions.
In the following we derive the equation of two-point  correlation function.

Adding the external
 source $J$ to  Eq.  \eqref{eqevpsiJ} and taking  the ensemble average,
we get
\bl \label{exppsieq}
&   \big \langle \ddot\psi \big \rangle_J  + 2H \big \langle \dot \psi \big \rangle_J
-\frac{c_s^2 }{a^2} \big \langle \nabla^2 \psi \big \rangle_J
- 4\pi G\rho_0  \big \langle (\psi^2 -\psi) \big \rangle_J
   \nn \\
&    - \frac{1}{a^2} \big \langle \nabla \psi \cdot\nabla \phi \big \rangle_J
   -\frac{1}{a^2} \big \langle \frac{\partial^2}{\partial x^i \partial x^j}(\psi  v^i v^j)
   \big \rangle_J
      -  \langle   J     \rangle_J  =0 .
\el
Applying functional derivative $\frac{\delta}{a^3 \beta\delta J(\textbf{x}^{\prime})}$
to each term in the above equation  and then setting  $J=0$,
we obtain Eq.\eqref{eq2ptcorr},
where the following have been  used.
The connected two-point correlation function of  the field $\psi$ is defined by
the ensemble average
\bl
G^{(2)}\left(\mathbf{x}_{1}, \mathbf{x}_{2}, t \right)
& \equiv\left\langle\delta \psi\left(\mathbf{x}_{1}, t  \right)
     \delta \psi\left(\mathbf{x}_{2},t \right)\right\rangle \nn \\
& = \frac{\delta }
   { ( a^3 \beta)  \delta J ( {\bf x}_{2})}
    \langle\psi\left(\mathbf{x}_{1}, t \right)  \rangle_{J}
     \Big |_{J=0} \nn \\
& =\left.\frac{\delta^{2}}{  ( a^3 \beta)^2 \delta J  ( {\bf x}_{2} )
     \delta J  ( {\bf x}_{1} )} \ln Z[J] \right|_{J=0}
     \label{defG2}
\el
where  $\delta \psi (\bf{x}) = \psi (\bf{x}) - \langle \psi (\bf{x}) \rangle$
is the dimensionless density fluctuation.
One has $\langle  \delta \psi({\bf r} ) \rangle =0$,
   $\langle \psi ({ \bf x }) \rangle|_{J=0}=1$,
and  $\delta \psi ({ \bf x}) = \delta ({ \bf x })$.
$G^{(2)}$ is dimensionless by the definition,
and is assumed to have the following  stationary property
\be \label{symmG2}
 G^{(2)}\left( { \bf x }_{1}, { \bf x}_{2}, t \right)
 = G^{(2)}\left( {\bf x}_{1} -{ \bf x}_{2}, t \right)
 =   G^{(2)} (| {\bf x}_1 - {\bf x}_2 |, t ) ,
\ee
which is consistent with the  isotropy of the background Universe.
The connected $n$-point correlation function is
\bl \label{Gndef}
G^{(n)}\left(\mathbf{r}_{1}, \ldots, \mathbf{r}_{n}\right)
& \equiv\left\langle\delta \psi\left(\mathbf{r}_{1}\right)
   \ldots \delta \psi\left(\mathbf{r}_{n}\right)\right\rangle \\
&=\left.   \frac{\delta^{n} \ln Z[J]}{ ( a^3 \beta)^{n} \delta J\left(\mathbf{r}_{1}\right)
       \ldots \delta J\left(\mathbf{r}_{n}\right)}\right|_{J=0} \\
&=\left. \frac{ \delta^{n-1}
   \left\langle\psi  (\mathbf{r}_{n} )
   \right\rangle_{J}}{ ( a^3 \beta)^{n-1} \delta J\left(\mathbf{r}_{1}\right)
     \ldots \delta J\left(\mathbf{r}_{n-1}\right)}\right|_{J=0},
     ~~~ \text{for $n\ge 2$} .
\el
Other terms are calculated   in the same manner,
\be\label{defG}
\frac{\delta}{ ( a^3 \beta) \delta J(\bf{x}^{\prime})}
\langle  \nabla^2 \psi({\bf x},t) \rangle \Big|_{J=0}
=  \nabla^2 G^{(2)}(\bf{x}-\bf{x}^{\prime}, t ), \nn
\ee
\be
\frac{\delta}{  ( a^3 \beta) \delta J(\textbf{x}')}
  \langle \dot\psi ({\bf x},t )  \rangle_J   \Big|_{J=0}
   = \dot  G^{(2)}({\bf x},{\bf x}',t), \nn
\ee
\be
\frac{\delta}{ ( a^3 \beta) \delta J(\textbf{x}')}
  \langle \ddot\psi({\bf x},t)  \rangle_J   \Big|_{J=0}
  = \ddot  G^{(2)}({\bf x},{\bf x}',t), \nn
\ee
\bl \label{3gd}
\frac{\delta}{ ( a^3 \beta) \delta J(\textbf{x}^{\prime}, t )}
\langle  \psi^{2}({\bf x})  \rangle \Big|_{J=0}
& = \frac{\delta}{a^3 \beta\delta J(\textbf{x}^{\prime})}
   \big(  \la \psi({\bf x}) \ra \la \psi({\bf x}) \ra
    +\la \delta  ({\bf x})  \delta ({\bf x}) \ra   \big)  \Big|_{J=0}
    \nn \\
& =2   G^{(2)}(\textbf{x}-\textbf{x}^{\prime}, t )
          +  G^{(3)}(\textbf{x},\textbf{x},\textbf{x}^{\prime} ; t ),
\el
where  $\langle \psi  \rangle_{J=0 } =1$ is used.
The external source term gives
\bl  \label{4b2}
\frac{\delta}{ (a^3 \beta )  \delta J(\textbf{x}')}
        \langle J({\bf x})   \rangle_J  \Big|_{J=0}
&   = \frac{1}{ a^3   \beta}    \delta^{(3)} (\bf{x}-\bf{x}')  .
\el
From these,  we get
\bl  \label{geq}
& \ddot  G^{(2)}({\bf x}-{\bf x}',t) + 2H \dot  G^{(2)}({\bf x}-{\bf x}',t)
  - \frac{c_s^2 }{a^2} \nabla^2 G^{(2)}({\bf x}-{\bf x}',t)
  \nn \\
&   - 4\pi G\rho_0 (t) \left(   G^{(2)}(\textbf{x}-\textbf{x}^{\prime},t )
          +  G^{(3)}(\textbf{x},\textbf{x},\textbf{x}^{\prime}, t ) \right)
           \nonumber \\
&   -   \frac{1}{a^2} \frac{\delta}{ ( a^3 \beta)\delta J(\textbf{x}^{\prime})}
           \langle \nabla \psi (\textbf{x}) \cdot\nabla \phi (\textbf{x}) \rangle \Big|_{J=0}
 -  \frac{1}{a^2} \frac{\partial^2}{\partial x^i \partial x^j}
      \frac{\delta}{ ( a^3 \beta)\delta J(\textbf{x}^{\prime})}
                 \langle (\psi (\textbf{x}) v^i (\textbf{x}) v^j (\textbf{x})) \rangle \Big|_{J=0} \nn \\
&   =  \frac{1}{a^3 \beta}    \delta^{(3)}(\bf{x}-\bf{x}') .
\el

To deal with   the potential term $\nabla \psi  \cdot\nabla \phi$ in \eqref{geq},
 we use the solution   of the Poisson equation \eqref{Poissonphi}
\be \label{solphi}
\phi({\bf x}, t)= -a^2 G \rho_0(t) \int \frac{
     \psi( {\bf x} ', t )- 1}{|\bf x-x'|}d^3 {\bf x}' ,
\ee
\bl
\nabla  \phi ({\bf x} ,t)
 = -a^2 G \rho_0(t)
     \int  \big(\psi({\bf x'} ,t) -1 \big )
        \nabla  \frac{1}{|\bf x-x'|}d^3 {\bf x}' \,  ,    \label{nblph}
\el
where $\nabla \equiv \nabla_x$.
So  we have
\be\label{rhophit}
\nabla \psi  \cdot\nabla \phi= -a^2 G \rho_0 (t)
   \int  \bigg(  \psi({\bf x'} ,t)
   \nabla \psi({\bf x} ,t) -\nabla \psi({\bf x} ,t) \bigg)
     \cdot  \nabla \frac{1}{|\bf x-x'|}d^3 {\bf x}'.
\ee
Applying functional differentiation
on the ensemble average of \eqref{rhophit}, we get
\bl  \label{fctdr}
-\frac{\delta}{a^3  \beta\delta J(\textbf{x}^{'})}
  \frac{1}{a^2}  \langle \nabla \psi  \cdot\nabla \phi \rangle
  \Big|_{J=0}
 = &   4\pi  G   \rho_0(t)  G^{(3)}(\textbf{x},\textbf{x},\textbf{x}^{'})
                 \nn \\
   &  +  G \rho_0 (t)
   \int  \nabla \cdot \Big(  G^{(3)}(\textbf{x},\textbf{x}',\textbf{x}^{''})
       \cdot  \nabla  \frac{1}{|\bf x-x''|}    \Big)  d^3{\bf x''}     .
\el
Substituting \eqref{fctdr} into  \eqref{geq},
noting that $4\pi G \rho_0  G^{(3)}$ in Eq.\eqref{fctdr}
will cancel the  $- 4\pi G \rho_0 G^{(3)}$ term in Eq.\eqref{geq},
we obtain Eq.  \eqref{eq2ptcorr} of two-point  correlation function.

The velocity dispersion term $\psi v^i v^j$ in \eqref{eq2ptcorr}
can be treated as the following.
Under the Zel'dovich approximation \cite{Peebles1980},
the peculiar  velocity $v^i$ can be expressed
in terms of the density field $\psi$
\be \label{Zeldovichapprox}
\mathbf{v}=-\frac{1}{4 \pi G \rho_{0} a} \frac{\dot{D}}{D} \nabla \phi
=\frac{H f\left(\Omega_{m}\right)}{4 \pi} a \int
      (\psi ({ \bf x}^{\prime}, t ) -1)
        \nabla       \frac{1}{\left|\mathbf{x}^{\prime}-\mathbf{x}\right|}
        d^{3} x^{\prime} + O (\delta^{2} )
\ee
where
\bl \label{fOmega}
f(\Omega_m,t ) \equiv -\frac{d \ln D}{d \ln (1+z)}
  =\frac{a}{D}\frac{d D}{d a },
\el
and  $D(t)$ is a growing mode of the linear part of  equation (\ref{nonlindelta})
without pressure  \cite{Peebles1980}.
For a flat RW spacetime,
it can be  approximately  fitted by the following formula \cite{Lahav1991},
\bl
f(\Omega_m,t ) =\Omega^{0.6} + \frac{1}{70} \big( 1-\frac12 \Omega (1+\Omega) \big),
      ~~ \text{with} ~~
\Omega \equiv \Omega_m \frac{(1+z)^3}{(\Omega_m (1+z)^3+ \Omega_\Lambda )^2 } .
\el
Substituting \eqref{Zeldovichapprox} into the velocity dispersion term
of \eqref{geq},
we calculate
\bl
& \frac{\delta}{a^3 \beta \delta J( {\bf x}')}
 \Big[  \la \psi({\bf x})\ra   \la\delta({\bf y})\delta({\bf z}) \ra
+  \la   \delta({\bf x}) \delta({\bf y})\delta({\bf z}) \ra \Big] \Big|_{J=0}
\nn \\
& =  G^{(2)}({\bf x-x'}) G^{(2)}({\bf y-z})
  +   G^{(3)}({\bf y, z, x'})
  +  G^{(4)}({\bf x},{\bf y},{\bf z},{ \bf x}') ,
\el
where  the definition \eqref{Gndef} has  been used,
and get   the following
\bl   \label{psivvyz}
&   \frac{\delta}{ a^3 \beta \delta J( {\bf x}')}
    \la \psi  v^i v^j \ra \Big|_{J=0}
 = \frac{H^2 a^2 f^2(\Omega_m)}{16 \pi^2}
  \iint d^3 y \, d^3 z
   \frac{y^i-x^i}{| {\bf y}-{\bf x}| ^3 }
    \frac{z^j-x^j}{|{\bf z}-{\bf x} | ^3 }
\nn \\
& \times \bigg(   G^{(2)}({\bf x-x'},t)G^{(2)}({\bf y-z},t)
  +  G^{(3)}({\bf y, z, x'};t)  + G^{(4)}({\bf x, y, z, x'};t) \bigg ) .
\el
Substituting \eqref{psivvyz}  into Eq.\eqref{eq2ptcorr}
yields  the nonlinear equation   \eqref{eq2ptcorr1}
which contains $G^{(3)}$ and $G^{(4)}$.
By comparison, Davies and  Peebles  \cite{DaviesPeebles1977}
did not use the Zel'dovich approximation,
so that their  Eq.(72)  contains the unknown velocity dispersions.
The Zel'dovich approximation will  cause an error
of  the order  $\delta^3$ in $v^i v^j$,
which would bring about
extra terms like  $G^{(4)}$  and  $G^{(2)} G^{(3)}$ in \eqref{psivvyz}.
We shall drop these terms.
Therefore, \eqref{psivvyz} is accurate up to
a numerical factor of the term $G^{(4)}$.

To make  Eq.  (\ref{eq2ptcorr1}) closed  for $G^{(2)}$,
we adopt the Kirkwood-Groth-Peebles ansatz
on the three-point  correlation function \cite{Kirkwood1932,GrothPeebles1977},
\bl\label{GrothPeeblesAnsatz}
G^{(3)}_{123}
  = Q \big[ G^{(2)}_{12} G^{(2)}_{23}
    + G^{(2)}_{23} G^{(2)}_{31}
    + G^{(2)}_{31} G^{(2)}_{12} \big],
    ~~~ Q=1.0\pm 0.2 ,
\el
where   $G^{(2)}_{12}$ denotes $G^{(2)}(\mathbf{x_1, x_2},t)$
and $G^{(3)}_{123}$ denotes $G^{(3)}(\mathbf{x_1, x_2, x_3};t)$
for  notational simplicity,
and $Q$ is a dimensionless constant to be determined by observation.
[We remark that
the ansatz \eqref{GrothPeeblesAnsatz} with $Q=1$
holds exactly as a solution of $G^{(3)}_{123}$
in the Gaussian approximation  for the static case \cite{ZhangMiao2009}.]
For the four-point  correlation function,
we adopt the Fry-Peebles ansatz \cite{FryPeebles1978}
\bl  \label{FryPeeblesAnsatz}
G^{(4)}_{1234}
= & R_a \Big[ G^{(2)}_{12} G^{(2)}_{23} G^{(2)}_{34}
            + G^{(2)}_{23} G^{(2)}_{34} G^{(2)}_{41}
            + G^{(2)}_{34} G^{(2)}_{41} G^{(2)}_{12}
             + G^{(2)}_{41} G^{(2)}_{12} G^{(2)}_{23} \nn \\
&  +  G^{(2)}_{13}  G^{(2)}_{34}  G^{(2)}_{42}
   +  G^{(2)}_{34}  G^{(2)}_{42}  G^{(2)}_{21}
   +  G^{(2)}_{42}  G^{(2)}_{23}  G^{(2)}_{31}
   +  G^{(2)}_{21}  G^{(2)}_{13}  G^{(2)}_{34}
\nn \\
&   +   G^{(2)}_{14} G^{(2)}_{42} G^{(2)}_{23}
    +   G^{(2)}_{31} G^{(2)}_{14} G^{(2)}_{42}
    +   G^{(2)}_{31} G^{(2)}_{12} G^{(2)}_{24}
    +  G^{(2)}_{23}  G^{(2)}_{31}  G^{(2)}_{14}  \Big] \nonumber \\
 + & R_b  \Big[ G^{(2)}_{12} G^{(2)}_{13} G^{(2)}_{14}
    + G^{(2)}_{21} G^{(2)}_{23} G^{(2)}_{24}
    + G^{(2)}_{31} G^{(2)}_{32} G^{(2)}_{34}
    + G^{(2)}_{41} G^{(2)}_{42} G^{(2)}_{43} \Big].
\el
where $G^{(4)}_{1234}$ denotes $G^{(4)}(\mathbf{x_1, x_2, x_3, x_4};t)$,
$R_a$ and $R_b$ are two parameters
and observations indicate  $3R_a+R_b \simeq 10 \pm 2$
\{see  (19.23)   in ref.\cite{Peebles1993}\}.
The undetermined numerical factor of  $G^{(4)}$ due to
the Zel'dovich approximation can be absorbed into the parameters $R_a$ and $R_b$.
Substituting \eqref{GrothPeeblesAnsatz} and  \eqref{FryPeeblesAnsatz}
into \eqref{eq2ptcorr1},
using the conventional notation $\xi({\bf x},t)=G^{(2)} ({\bf x},t)$,
we arrive at the nonlinear  Eq.  \eqref{eq2ptcorr34}
in the context.

By similar calculations,
taking functional derivative of the ensemble average of
the continuity equation \eqref{continuityexpan} and
using the Zel'dovich approximation \eqref{Zeldovichapprox},
we obtain the continuity equation in terms of correlation function
\be\label{cont2pgr2}
 \frac{\partial }{\partial t}\xi(\textbf{x}-\textbf{x}', t )
 -  H  f(\Omega_m)  \xi(\textbf{x} -\textbf{x}', t )
    + \frac{H  f(\Omega_m)}{4 \pi}
  \int  \nabla  \cdot \Big(   G^{(3)}(\textbf{x}, \textbf{x}^{'},\textbf{x}^{''})
  \nabla \frac{1}{\lvert {\bf x''}-{\bf x} \rvert} \Big) d^3 x'' =0 .
\ee
This is closed when  the Groth-Peebles ansatz \eqref{GrothPeeblesAnsatz} is used.
Equation \eqref{cont2pgr2} should be compared with
the conservation of particle  pair  \{Eq. (41) or  Eq. (71b) in Ref.\cite{DaviesPeebles1977}\}
which still  contains the relative velocity of a pair of galaxies.
We have made use of  \eqref{cont2pgr2} to give
an estimate  of  the change rate \eqref{raterv} in the context.

\section{The linear solution in terms of the Green's function}

Although we have obtained the numerical linear solution $P_k$ and  $\xi$,
it is revealing  to analyze
the solution   by the Green's function method.
To isolate the influences of the source $A$ and  the initial conditions,
in accordance  with   \eqref{xi1xi2} we write
\bl \label{splitw1w2}
\xi({\bf x}, t )= \frac{w({\bf x}, t )}{a(t)}
= \frac{1}{a}(w_1({\bf x}, t )+w_2({\bf x}, t )) ,
\el
where  $w_1$
satisfies   the following
\be \label{vHelmholtzeq}
\left\{
\begin{split}
&   \ddot w_1  ({\bf x},t)
  - c^2(t) \nabla^2  w_1 ({\bf x},t)
  -  m^2(t)  w_1 ({\bf x},t)
  =   \frac{A}{ a(t)^2 } \delta^{(3)}({\bf x})  \\
&  w_1|_{t=t_i } = 0 \\
& \dot  w_1 \Big|_{t=t_i}
    = 0   ,\\
\end{split}
\right.
\ee
with
\bl
c^2(t) &  \equiv   \frac{c_{s0}^2}{ a^{2+2\eta}(t)},
\\
m^2(t ) & \equiv 4 \pi G \rho_c
           \Big[\frac23 ( - \frac12 a^{-3}\Omega_m + \Omega_\Lambda )
   + \frac{ \gamma \,  \Omega_m }{a^{3}(t)} \Big]  , \label{coe}
\el
where  the Friedmann equation
$\frac{\ddot a}{a} = \frac{8\pi G \rho_c }{3}
( - \frac12 a^{-3}\Omega_m + \Omega_\Lambda )$  has been used.
$w_1$  is contributed by  the source $A$.
In \eqref{splitw1w2},
$w_2$   satisfies  the following
\be  \label{uHelmholtzeq}
\left\{
\begin{split}
&   \ddot w_2  ({\bf x},t)
   - c^2(t)  \nabla^2  w_2 ({\bf x},t)
  -  m^2(t)  w_2 ({\bf x},t)
  =  0 , \\
& w_2|_{t=t_i } = a(t_i) \xi({\bf x}, t_i)    ,\\
& \dot  w_2 \big|_{t=t_i}
      =a(t_i)  \Big( H(t_i) \xi({\bf x}, t_i)
          + \dot\xi({\bf x}, t_i) \Big)   ,       \\
\end{split}
\right.
\ee
where the initial time $t_i$ corresponds to the  redshift $z=7$,
and  the  $\xi({\bf x}, t_i)$ and  $\dot \xi({\bf x}, t_i)$
are given by the initial spectrum $P_{k \, ini }$ of  \eqref{iniPkold}
and  the initial  rate $r_a$ of \eqref{raterv}.
$w_2$  is contributed by  the initial condition.
Corresponding to \eqref{splitw1w2},
the  power spectrum is also split into  two parts
\be \label{decompsolhoinho}
P_k   =  \frac{1}{a} \int d^3 x \, \Big( w_1({\bf x}, t)+ w_2({\bf x}, t) \Big)
        e^{-i{\bf k \cdot x}}
   =  P_{k\, 1}  + P_{k\, 2} .
\ee
We  obtain numerically the solutions $P_{k\, 1}$ and $\xi_1$
shown in   Fig.\ref{16} and  Fig.\ref{17},
as well as $P_{k\, 2}$ and $\xi_2$
 in  Fig.\ref{18},  Fig.\ref{19}, and Fig.\ref{20}.

\begin{figure}
\centering
\includegraphics[width=0.5 \textwidth]{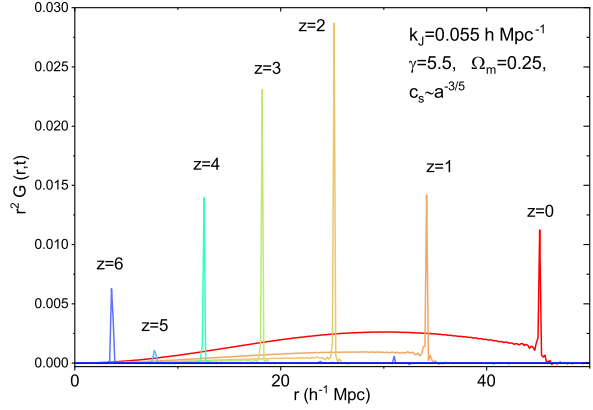}
\caption{The Green's function
$G(r,t)$ as the solution of \eqref{homoequG} is a spike at each instant,
analogous to a delta function $\delta(r-c(t)t)$,
and is propagating forward at a speed $ c(t)$.
The weighted  $r^2 G(r,t)$ is shown. }
\label{25}
\end{figure}

\begin{figure}
\centering
\includegraphics[width=0.5 \textwidth]{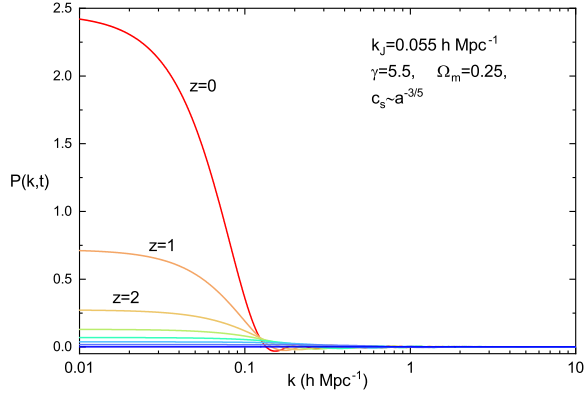}
\caption{Same as Fig. \ref{25}.
The power spectrum $P_k(t)$ associated with $G(r,t)$.}
\label{26}
\end{figure}

We now express these the solutions in terms of the Green's function.
It can be checked  that the principle of homogeneity
 also applies to the inhomogeneous equation \eqref{vHelmholtzeq}
with the  time-dependent coefficients.
So, by  Duhamel's principle,
Eq.  \eqref{vHelmholtzeq} of $w_1$
reduces to a homogeneous equation with a  nonzero initial velocity,
the solution is given by
\bl
w_1  ({\bf x},t)
& =  \int^t_{t_i} \Big[\int   \frac{A}{ a^2 (\tau) } \delta^{(3)}({\bf x_0})
              G({\bf x,x_0}, t) d^3 x_0 \Big] d\tau
           \nn \\
& =  \int^t_{t_i} \frac{A}{ a^2 (\tau)} G({\bf x}, t) d\tau   ,
          \label{inhmhom2}
\el
where  $G({\bf x,x_0}, t)$  is the Green's function satisfying  the following
\be \label{homoequG}
\left\{
\begin{split}
& \ddot G ({\bf x, x_0},t) - c^2(t) \nabla^2  u ({\bf x ,x_0 },t)
   - m^2(t) G ({\bf x ,x_0},t)
          =  0        ~~~ ( t> t_i  )
    \\
& G  \big|_{t=t_i } = 0
   \\
& \dot  G \big|_{t=t_i }  = \delta^{(3)}({\bf x-x_0})  ,   \\
\end{split}
\right.
\ee
which   describes the field at $\bf x$
that is generated by a point source located at $\bf x_0$,
and is propagating at a  speed $c(t)$.
In the simple case $c(t)=c_s$ and $m(t)=0$,  one would get
the well-known expression
\bl  \label{Greens}
G ({\bf x, x_0},t) = \frac{1}{4\pi c}
   \frac{\delta(|{\bf x-x_0}|-c_s t )}{|\bf x-x_0|}  ,
\el
which is the field  propagating at a speed $c_s$.
In our case with the time-dependent coefficients $c(t)$ and $m(t)$,
the analytical expression of $G({\bf x, x_0},t)$ is hard to get.
Nevertheless,  the numerical solution is obtained,
and shows a  behavior  analogous to \eqref{Greens},
as plotted in Fig.\ref{25} and Fig.\ref{26}.
One sees that  $G({\bf x, x_0},t)$
is a sharp spike  located at $({\bf x-x_0})$ at instance $t$,
like  $\delta(|{\bf x-x_0}|-c(t) t)$,
and is propagating forward at finite speed $c(t)$,
which is quite  similar to   \eqref{Greens} in the simple case.
After the $d\tau$ integration in \eqref{inhmhom2},
$ w_1$ receives contributions from all the spherical surfaces of the radius $\leq c(t)t$,
analogous to a step function,
nonvanishing only within a region $r<c(t)t$.
The inhomogeneous solution $\xi_1=w_1/a$ is the major part of
the main mountain of $\xi$ at small scales,
and  is growing with time.
This  ``retarded potential" behavior of $\xi_1$
is  demonstrated in   Fig.\ref{16}.

The solution $w_2$ of \eqref{uHelmholtzeq}  also can be expressed
in terms of Green's function.
We  decompose it into two parts,
\bl \label{w2uv}
w_2({\bf x}, t)  =  u({\bf x}, t) +  v({\bf x}, t) ,
\el
where    $u$ satisfies  the following homogeneous equation with the initial velocity
\be \label{homoequ2}
\left\{
\begin{split}
& \ddot u ({\bf x},t  ) - c^2(t) \nabla^2  u ({\bf x},t )
  - m^2(t) u ({\bf x},t )       =  0
         \\
&   u \big|_{t=t_i} = 0 \\
&   \dot  u \big|_{t=t_i}
  = a(t_i)  \Big( H(t_i) \xi({\bf x}, t_i)
      + \dot\xi({\bf x}, t_i) \Big)  ,
\end{split}
\right.
\ee
which has the same structure as  \eqref{homoequG}
and  the solution  is
\bl \label{uinhom}
 u({\bf x}, t )
 = \int        a(t_i)   \Big( H(t_i) \xi({\bf x_0},t_i)
        + \dot\xi({\bf x_0}, t_i) \Big)
         G({\bf x,x_0}, t) d ^3 x_0    ,
\el
where $G({\bf x,x_0}, t)$ is the  Green's function  of \eqref{homoequG}.
$u$ represents the contribution of the initial rate.
Due to the property of  $G({\bf x,x_0}, t)$,
the $d ^3 x_0 $ integration receives  most of  contributions
 from a region around a spherical surface  of
 radius $c(t) t$ centered at ${\bf x}$.
$v$   in \eqref{w2uv}   satisfies the following equation with the initial value
\be \label{homoeqv}
\left\{
\begin{split}
& \ddot v ({\bf x},t) - c^2(t) \nabla^2  v ({\bf x},t)
  - m^2(t) v ({\bf x},t)       =  0         \\
& v({\bf x},t)|_{t=t_i } =  a(t_i)  \xi({\bf x}, t_i)    \\
& \dot  v ({\bf x},t)  \big|_{t=t_i }  = 0 .     \\
\end{split}
\right.
\ee
By Duhamel's principle,    the solution  is
\bl \label{vhom}
v({\bf x},t) =  v({\bf x} , t_i)
+\int^{t} _{t_{i}}
  \Big[   \int    \Big(c^2(\tau ) \nabla^2_{\bf x_0}   v({\bf x_0}, t_i)
       + m^2(\tau )v({\bf x_0}, t_i) \Big) G({\bf x, x_0}, t) d^3 x_0  \Big]   d \tau ,
\el
where $G({\bf x,x_0}, t)$ is the Greens' function of \eqref{homoequG}.
The homogeneous solution $\xi_2 =(u+v)/a$
gives rise to the periodic bumps of $\xi$,
whose amplitudes are growing during evolution,
as seen in Fig.\ref{18}, Fig.\ref{19},  and Fig.\ref{20}.
$v$ is dominant over $u$ because the rates $H_0$ and $r_a$ are small.
The sum of   \eqref{inhmhom2},  \eqref{uinhom}, and   \eqref{vhom}
is  the  solution $\xi$ of  Eq. \eqref{linapprgtcs0}
in terms of the  Green's function.

\end{document}